\pdfoutput=1 

\documentclass[12pt,usenames,dvipsnames,a4paper]{article}

\usepackage{amsmath,amsfonts,amssymb}
\usepackage{appendix}
\usepackage{color}
\usepackage{datetime}
\usepackage{graphicx}
\usepackage{dsfont}
\usepackage{array}
\usepackage[citecolor=blue]{hyperref}

\usepackage{subfig}
\usepackage[compat=1.1.0]{tikz-feynman} 

\def\a{\alpha}

\def\b{\beta}
\def\c{\chi}

\def\d{\delta}
\def\D{\Delta}

\def\eps{\varepsilon}
\def\f{\frac}
\def\g{\gamma}

\def\G{\Gamma}

\def\l{\left}

\def\la{\langle}
\def\ra{\rangle}

\def\mc{\mathcal}

\def\m{\mu}
\def\n{\nu}

\def\nn{\nonumber}

\def\p{\partial}

\def\r{\right}

\def\dst{\displaystyle}

\def\be{\begin{equation}}
\def\ee{\end{equation}}

\def\bea{\begin{eqnarray}}
\def\eea{\end{eqnarray}}

\def\ba{\begin{array}}
\def\ea{\end{array}}

\def\bc{\begin{center}}
\def\ec{\end{center}}

\def\bl{\begin{flushleft}}
\def\el{\end{flushleft}}

\def\br{\begin{flushright}}
\def\er{\end{flushright}}

\def\bi{\begin{itemize}}
\def\ei{\end{itemize}}

\def\bt{\begin{tabular}}
\def\et{\end{tabular}}

\def\dd{\mathrm d}

\newcolumntype{P}[1]{>{\centering\arraybackslash}p{#1}}
\newcolumntype{M}[1]{>{\centering\arraybackslash}m{#1}}

\newcommand{\mO}{\mathcal{O}}

\newcommand{\btf}[1]{\begin{tikzpicture}[baseline=-3pt] \begin{feynman}[inline=(#1.base)]}
\def\etf{\end{feynman}\end{tikzpicture}}

\tikzfeynmanset{
  bigdot/.style = {
    /tikzfeynman/dot,
    /tikz/minimum size=4pt,
  },
  bigcross/.style = {
    /tikzfeynman/crossed dot,
    /tikz/minimum size=6pt,
  }
}

\begin{document}

\thispagestyle{empty}

\begin{center}
\bf \Large{The Epsilon Expansion Meets Semiclassics}
\end{center}

\begin{center}
{\textsc {Gil~Badel$^1$, Gabriel~Cuomo$^1$, Alexander~Monin$^{1,2}$, Riccardo~Rattazzi$^1$}}
\end{center}

\begin{center}
{\it $^1$Institute of Physics, Theoretical Particle Physics Laboratory (LPTP), \\ 
\'Ecole Polytechnique F\'ed\'erale de Lausanne (EPFL), \\ 
CH-1015 Lausanne, Switzerland}

\vspace{0.5cm}
{\it $^2$Department of Theoretical Physics, University of Geneva, \\
24 quai Ernest-Ansermet, 1211 Geneva, Switzerland}
\end{center}

\begin{center}

\texttt{\small gil.badel@epfl.ch}  \\
\texttt{\small gabriel.cuomo@epfl.ch}  \\
\texttt{\small alexander.monin@unige.ch}  \\
\texttt{\small riccardo.rattazzi@epfl.ch} \\

\end{center}

\vspace{2cm}

\abstract{We study the scaling  dimension $\Delta_{\phi^n}$ of the operator $\phi^n$ where $\phi$ is the fundamental complex field of the $U(1)$ model at the Wilson-Fisher fixed point in $d=4-\varepsilon$. 
Even for a perturbatively small fixed point coupling $\lambda_*$, standard perturbation theory breaks down for sufficiently large $\lambda_*n$. Treating $\lambda_* n$ as fixed for small $\lambda_*$ we show that $\Delta_{\phi^n}$ can be successfully computed through a semiclassical expansion around a non-trivial trajectory, resulting in
\be\nonumber
\Delta_{\phi^n}=\frac{1}{\lambda_*}\Delta_{-1}(\lambda_* n)+\Delta_{0}(\lambda_* n)+\lambda_* \Delta_{1}(\lambda_* n)+\ldots
\ee
We explicitly compute the first two orders in the expansion, $\Delta_{-1}(\lambda_* n)$ and $\Delta_{0}(\lambda_* n)$. The result, when expanded at small $\lambda_* n$, perfectly agrees with all available diagrammatic computations. The asymptotic at large $\lambda_* n$ reproduces instead the systematic large charge expansion, recently derived in CFT. Comparison with Monte Carlo simulations in $d=3$ is compatible with the obvious limitations of taking $\varepsilon=1$, but encouraging.}

\newpage{}

\tableofcontents

\section{Introduction}

Quantum Mechanics is  an astonishing fact of Nature, but a moment's thought identifies the origin of the astonishment: the amazing ability of quantum mechanics to disappear behind classical physics in a vast array of physical situations. Those of course  include life on earth,  which molded our mind  in the course of  an odd billion years of evolution.

The different regimes of quantum systems are readily classified through the properties of the corresponding path integral. The latter of course depends not only on the dynamics  but also on the boundary conditions and, somewhat equivalently, on the operator insertions. Path integrals can be  broadly divided  into two classes, Weakly Coupled (WC) and Strongly Coupled (SC). A path integral is weakly coupled when it can be approximated by a loop expansion around some leading classical trajectory $\gamma$.
In the WC case the contribution to physical observables ${\cal O}$ consists of the sum of two terms ${\cal O}={\cal O}_\gamma+ {\cal O}_q$, a classical one ${\cal O}_\gamma$, determined by the value of physical variables along the leading trajectory, and a quantum one, ${\cal O}_q$, determined by quantum fluctuations around $\gamma$. One can then distinguish classical and quantum observables depending respectively on whether  ${\cal O}_\gamma\gg {\cal O}_q$ or not. For instance, the path integral for the harmonic oscillator is always weakly coupled, by definition,  while the corresponding ground state energy is a quantum observable, and  one can consider states (for instance coherent states) where to find classical observables. A strongly coupled path integral instead occurs when no saddle point approximation is possible. In that case all observables are quantum mechanical. An example of this situation is given by QCD processes around the GeV. In fact QCD suggests also  a third class, given by strongly coupled path integrals  where the quantum fluctuations of at least a  subset of the variables, normally associated to long distance physics, are small around some trajectory. In that case one can first integrate out the variables with large quantum fluctuations and derive an effective weakly coupled description for the remaining variables. In the case of QCD, the latter correspond to the low energy excitations of the pions.

The most common practice in particle physics concerns processes involving a few weakly interacting particles, for instance $1\to 2$, $2\to 2$, $2\to 3 $, etc. That corresponds to computing quantum fluctuations around the vacuum trajectory in a weakly coupled path integral. On the other hand it is well known that, even in weakly coupled QFT, when considering processes whose number of legs $n$ grows, perturbation theory eventually fails \cite{Rubakov:1995hq}.
This issue was investigated in some details in the 80's and 90's, where, focussing on massive $\lambda\phi^4$, some remarkable results were obtained. In particular, it was shown that the computation could be organized as a semiclassical expansion around a non-trivial trajectory \cite{Son:1995wz}.
Mostly technical  difficulties, but also some conceptual ones, however slowed  progress down. A recent revival \cite{Khoze:2018mey} did not greatly progress, in our opinion, towards the tackling of the difficulties (see for instance \cite{Monin:2018cbi,Belyaev:2018mtd} for a critical assessment). This remains an important problem, not only technically and conceptually, but also phenomenologically, when considering the fate of scattering amplitudes involving many $W,Z$ and Higgs bosons in the Standard Model (SM) at energies that may be approachable at the next generation of colliders. While keeping the fate of the SM  in our mind, in the present paper we shall instead focus on a simpler problem, plausibly the simplest one in the context of multilegged amplitudes. We shall study the correlator of the operator of charge $n$, $\phi^n$, in  $U(1)$ invariant scalar QFT with quartic interaction $(\bar \phi\phi)^2$. In particular we shall study its scaling dimension, mostly focussing on the Wilson-Fisher fixed point in $d=4-\varepsilon$, at small $\varepsilon$ where the coupling is weak. The main conceptual result of our paper is that the operator's scaling dimension $\Delta_{\phi^n}$ can be computed through a systematic expansion around a non-trivial  trajectory, yielding
\be
\label{first}
\Delta_{\phi^n}=\frac{1}{\lambda_*}\Delta_{-1}(\lambda_* n)
+\Delta_{0}(\lambda_* n)+\lambda_*\Delta_{1}(\lambda_* n)+\dots
\ee
with $\lambda_*/16\pi^2=\varepsilon/5+\dots$  the fixed point coupling, and with $\D_{\ell-1}$ representing the $\ell$-th loop contribution. This result will be made concrete through the explicit computation of the leading and subleading terms, $\D_{-1}$ and $\D_0$.
Eq.~(\ref{first})  shows the existence of a double scaling limit, where $\lambda_*\to 0$ and $n\to \infty$ with $\lambda_* n$ fixed, where $\lambda_*$ remains the loop expansion parameter, while the effects of large $n$ are controlled by the classical parameter $\lambda_* n$. Our system, when weakly coupled around the vacuum, thus remains weakly coupled also at large $n$.
However our result applies equally well to large  and to small $\lambda_* n$, where one can also compute using Feynman diagrams. On the one hand this illustrates that the poor behaviour of standard perturbation theory as $\lambda_* n$ is  increased is simply tied to a poor choice of the path integral trajectory around which to expand. On the other hand it allows to compare our semiclassical computation to the results obtained using Feynman diagrams. In that doing we shall not only find perfect agreement, but  also be able to combine our result with finite order calculations and predict expansion coefficients that are beyond the order reached by each method when taken individually.
 
The simplicity  of the problem we consider, we believe, illuminates previous literature in related but different contexts. As concerns multilegged scattering amplitude, the structure of our computation is precisely the same, and precisely identical is the emergence of a double scaling limit, $\lambda\to 0$ with $\lambda n$ fixed. This indicates a sort of universality in the structure of multilegged observables, with $\lambda n$ acting like a sort of '{\nolinebreak}t Hooft coupling, and motivates further investigations into the more difficult problem of particle production. On the CFT side, our result directly connects to recent work on the general properties of large charge operators \cite{Hellerman:2015nra,Monin:2016jmo,Grassi:2019txd}. In that context,  it shows more concretely how the superfluid configuration of the leading trajectory emerges and it offers a concrete ``UV" complete realization of the effective field theory describing the superfluid. In particular the parameter $\lambda_* n$  controls the occurrence of the pure superfluid regime: at small $\lambda_*n$ the leading trajectory corresponds to a superfluid interacting with a light radial excitation, while at large $\lambda_*n$ the latter decouples. In our amusingly simple scenario, the parameter $\lambda_* n$ thus seems to play a role similar to the 't Hooft coupling in AdS/CFT, where it controls the gap between stringy and supergravity modes.
Finally, our systematic expansion in $\varepsilon$ invites a comparison with the results of Monte Carlo simulations in
$d=3$. While we are aware that taking $\varepsilon = 1$ is a significant stunt, we nonetheless find the comparison encouraging already with the first two orders we computed. This warrants computation of the next order, $\Delta_1$. 

This paper is organized as follows. In section \ref{secPTvacuum} we setup our conventions and we review the standard perturbative calculation of the anomalous dimension of $\phi^n$. In section \ref{secSemiclassical} we derive the existence of the expansion \eqref{first} and we show how to compute the leading term $\Delta_{-1}$ for small $\lambda_* n$ within the proposed approach. Section \ref{secCylinder} deals with the explicit calculation of the first two leading terms in \eqref{first} for arbitrary values of $\lambda_* n$; the result is analyzed at length in section \ref{secDiscussion}. We finally comment on future directions in \ref{secOutlook}.

\section{Perturbation theory around the vacuum}\label{secPTvacuum}

\subsection{Conventions}

In this  paper we will consider massless  $U(1)$ symmetric $\lambda (\bar \phi \phi)^2$ theory  in $d=4-\eps$ dimensional euclidean space-time with lagrangian
\be
\mc L = \p \bar \phi \p \phi +\f{\lambda_0} {4} \l ( \bar \phi \phi\r )^2.
\label{eq:phi4Lagrangian}
\ee
We will first consider general coupling, but  we shall later derive more specific results by focussing on the Wilson-Fisher fixed point. Renormalized field and coupling are defined according to
\be
\phi = Z_\phi [\phi], ~~ \lambda_0 = M ^ {\eps} \lambda Z_\lambda,
\label{eq:bareRenormalized}
\ee
where $M$ is the sliding scale. Throughout the paper we will adopt the minimal subtraction scheme, where
 $Z_\phi$ and $Z_\lambda$ are expressed as an ascending series of pure poles. In particular we have
\be
\log Z_\lambda = \sum_k \f {z_k(\lambda)}{\eps^k}=\f{c_{11}\lambda +c_{12}\lambda^2+\dots}{\eps}+ \f{c_{22}\lambda^2+\dots}{\eps^2}+\dots,
\ee
where 
\begin{equation}\label{eq:counterterm}
z_1(\lambda)=5\frac{\lambda}{(4\pi)^2}-\frac{15}{2}\frac{\lambda^2}{(4\pi)^4}+
\mO\l(\frac{\lambda^3}{(4\pi)^6}\r).
\end{equation}
Notice moreover that $Z_\phi=1$ up to two loop corrections. 
Using (\ref{eq:bareRenormalized}) one can easily show that the $\beta$-function equals
\be
\f{\p \lambda}{\p\log M}\equiv\beta(\lambda)= -\eps\lambda + \beta_4(\lambda),
\ee
with
\be
\b_4(\lambda) = \lambda^2 \f{\p z_1}{\p \lambda}=5\frac{\lambda^2}{(4\pi)^2}
-15\frac{\lambda^3}{(4\pi)^4}+\mO\left(\frac{\lambda^4}{(4\pi)^6}\right).
\ee
At the Wilson-Fisher fixed point, defined by  $\lambda=\lambda_*$ such that $\beta(\lambda_*)=0$, the theory is invariant under conformal transformations. The fixed point coupling $\lambda_*$ is non-trivially 
determined by the space-time dimensionality
\begin{equation}\label{eq:lambda}
\frac{\lambda_*}{(4\pi)^2}=\frac{\eps }{5}+\frac{3 }{25}\eps^2+\mO(\eps^3).
\end{equation}
For $\eps\ll 1$ the theory is weakly coupled. As we will show in the next subsection, this does not prevent perturbation theory around the vacuum to break down for specific observables. 

\subsection{Anomalous dimension of large charge operators}

We will study the scaling dimension of the simplest  operator with $U(1)$ charge\footnote{In our conventions, $\phi,\bar{\phi}$ have charge, respectively, $1$ and $-1$.} $n$ (-$n$), denoted by  $[\phi^n]$ ($[\bar{\phi}^n]$) and related to the bare field by 
\begin{equation}\label{eq:renormalizedPhiN2}
\phi^n=Z_{\phi^n} [\phi^n]\, .
\end{equation}
where $Z_{\phi^n}$ is a multiplicative renormalization factor. The anomalous dimension is then given by
\be\label{eq:gammaPhiN}
\g_{\phi^n}=\f{\p \log Z_{\phi^n}}{\p \lambda} \, \l [ -\eps \lambda + \beta_4(\lambda) \r ]\,.
\ee
For arbitrary $\lambda$, $\g_{\phi^n}$ is scheme dependent and thus unphysical beyond leading order. That can easily be seen by changing the scheme according to $[\phi^n]\to f(\lambda)[\phi^n]$ and $Z_{\phi^n}\to Z_{\phi^n}/f(\lambda)$, with $f(\lambda)$ a power series with finite coefficients. In the new scheme the anomalous dimension is modified according to $\g_{\phi^n}\to \g_{\phi^n}-\beta(\partial_\lambda\ln f )$. On the other hand  $\beta(\lambda_*)=0$, so that  $\g_{\phi^n}$ is scheme independent and physical at the fixed point. Indeed, a straightforward solution of the Callan-Symanzik equation for $\langle [\bar \phi^n][\phi^n]\rangle$ shows that  the  operator's physical dimension at the fixed point is
\be
 \D_{\phi^n} =n (d/2-1)+\g_{\phi^n}(\lambda_*)\,.
 \ee
 
 \begin{figure}[t]
   \centering
   \includegraphics{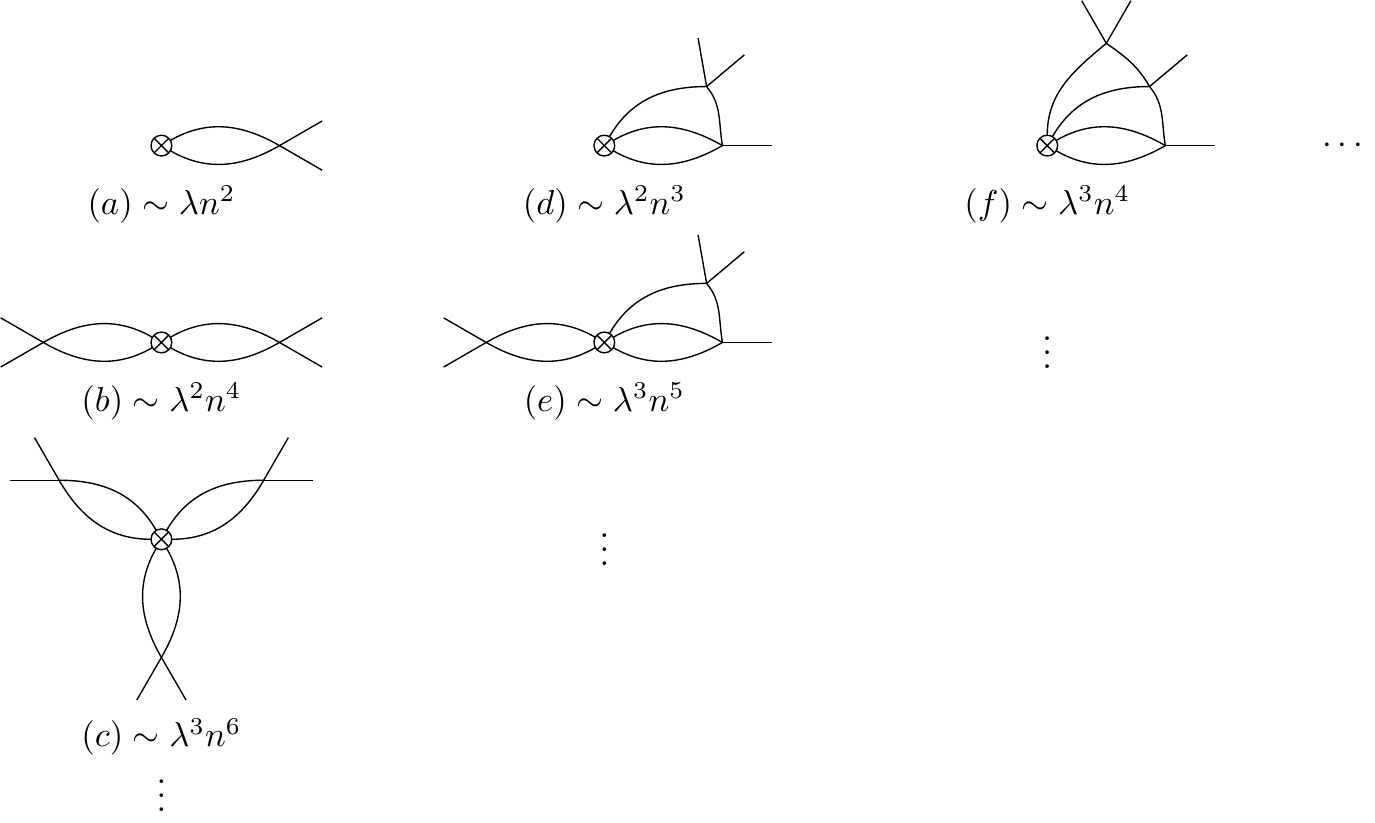}
   \caption{\label{fig:diagrammaticAnalysis} Some characteristic Feynman diagrams that appear with the $\phi^n$ operator.}
 \end{figure}

 We want to focus on $n\gg 1$, the regime of large charge or many legs. A first diagrammatic analysis shows multiplicity factors that grow with $n$, see figure \ref{fig:diagrammaticAnalysis}. Considering any loop order $\ell\ll n$, one finds contributions to  $Z_{\phi^n}$ that range from $\lambda^\ell n^{2\ell}$, for the daisy diagrams in the leftmost column of figure \ref{fig:diagrammaticAnalysis}, down to $\lambda^\ell n$, for corrections on single legs. In particular the ``connected diagrams", for which the number of legs picked from the $\phi^n$ equals $\ell+1$, like those in the top line of figure \ref{fig:diagrammaticAnalysis}, scale like $\lambda^\ell n^{\ell+1}$. However, a more detailed analysis shows that the terms with the highest powers of $n$ at any given loop order simply exponentiate terms from lower loops\footnote{As an illustration, it is simple to check that the sum over daisy diagrams exponentiate the $\lambda n^2$ contribution from the single petal diagram (a).}. As a consequence, in the expansion of  $\ln Z_{\phi^n}$, and thus of $\g_{\phi^n}$, the leading contribution at order $k$ scales like  the connected diagram, $\lambda^\ell n^{\ell+1}$. That is
 \be
 \g_{\phi^n}=n\sum_{\ell = 1} \lambda^\ell P_\ell(n)\,,
 \label{gammastructure}
 \ee
 with $P_\ell$ a polynomial of degree $\ell$. In truth we have explicitly checked that only up to four loops, but in the next section we shall give a general argument bypassing the diagrammatic analysis. The above result shows that, no matter how weakly coupled the theory is, for sufficiently large $\lambda n$, perturbation theory breaks down. The series in eq.~(\ref{gammastructure}) can also be organized in terms of leading and subleading $n$-powers, in close analogy with leading and subleading logs in the RG resummation
 \be
 \g_{\phi^n}=n\sum_{\kappa = 0} \lambda^\kappa F_\kappa(\lambda n)\,.
 \label{gammastructure2}
 \ee
 Very much like for the RG, this alternative rewriting of the series suggests an alternative loop expansion, performed after resumming (or straight out computing) all powers of $\lambda n$. Again, the physics underlying this alternative interpretation will be made manifest in the next subsections. Notice in passing, and consistently with the results in the next section, that the leading-$n$ contribution $F_0(\lambda n)$ is unaffected by changes in the subtraction scheme, like for instance $\lambda \to \lambda + a \lambda^2$ or $Z_{\phi^n}\to Z_{\phi^n}(1+b n^2\lambda)$, the latter corresponding to a simple reshuffling of the finite terms in the daisy diagram (a).

 Before moving  forward we would like to present the results of the explicit computation at 2-loops, whose details  are given in the appendix \ref{app:2loopPlane}. We shall need these  in order to compare to the 
 results of the more powerful method we shall develop in the next sections.
Working in the minimal subtraction scheme, we find 
\begin{equation}\label{eq:ZphiN}
Z_{\phi^n}=  1 - \frac{\lambda n (n-1)}{(16\pi^2) 2 \varepsilon} + \frac{\lambda^2}{(16\pi^2)^2} \left(\frac{n^4-2n^3-9n^2+10n}{8\varepsilon^2}+ \frac{2n^3-2n^2-n}{8\varepsilon}\right),
\end{equation}
which implies
\be
\g_{\phi^n} = n\l[\f{\lambda }{16\pi^2} \, \f{(n-1)}{2} - \l (\f{\lambda }{16\pi^2} \r)^2 \, \f{2n^2-2n-1}{4} \r]\,.
\label{eq:gammaTwoLoop}
\ee
Considering the theory at the fixed point this implies
\be
\D_{\phi^n} = n \l [ \l ( \f {d} {2} -1 \r )+\f{\varepsilon }{10} \, (n-1)- \f{\varepsilon^2 }{100} (2n^2-8n+5) \r ].
\label{eq:DeltaTwoLoop}
\ee

\section{Semiclassical approach}\label{secSemiclassical}

The scaling dimension of $[\phi^n]$ can also be directly computed by  considering
 the two-point function 
\be\label{eq:corr_lambda}
\la \bar{\phi}^n(x_f) \phi^n (x_i) \ra \equiv \f {\int \mc D \phi \mc D \bar \phi\, \bar \phi^n (x_f)\phi^n (x_i) \exp \l [ - \int {\cal L} \r ]} {\int \mc D \phi \mc D \bar \phi \, \exp \l [ -\int {\cal L} \r ]}\equiv Z_{\phi^n}^2 \la [\bar{\phi}^n](x_f) [\phi^n] (x_i) \ra\,.
\ee
The above integral can be cast in a form which exhibits its semiclassical nature in the small $\lambda$ regime independently of the size of  $n$. First it is convenient to rescale the field $\phi\to \phi/\sqrt{\lambda_0}$ to exhibit $\lambda_0$ as the loop counting parameter
\be
 \int {\cal L}\to \frac{1}{\lambda_0}\int \l [\p \bar \phi \p \phi +\f{1} {4} \l ( \bar \phi \phi\r )^2\r ]\equiv \frac{S}{\lambda_0}\,.
 \ee
Secondly $\bar{\phi}^n(x_f) \phi^n (x_i) $ can be brought up in the exponent, obtaining
\be\label{eq:corr_lambda2}
Z_{\phi^n}^2 \lambda_0^{n}\la [\bar{\phi}^n](x_f) [\phi^n] (x_i) \ra=\f {\int \mc D \phi \mc D \bar \phi\,  e^{-\frac{1}{\lambda_0}\l [\int \p \bar \phi \p \phi +\f{1} {4} \l ( \bar \phi \phi\r )^2-\lambda_0 n\left (\ln \bar \phi(x_f)+ \ln \phi(x_i)\right)\r ]}} {\int \mc D \phi \mc D \bar \phi \, e^{-\frac{1}{\lambda_0}\l [\int \p \bar \phi \p \phi +\f{1} {4} \l ( \bar \phi \phi\r )^2\r]}} \,.
\ee
The dependence on $\lambda_0$ and $n$, shows that we can perform the path integral using a saddle point expansion in the limit of small $\lambda_0$, while keeping $\lambda_0 n$ fixed. This limit thus encompasses 
the case where $\lambda_0 n$ is (arbitrarily) large\footnote{Of course we are making here a formal statement by using the bare coupling,
which is a power series in the renormalized coupling. In terms of renormalized quantities the limit is thus $\lambda(M)$ small with $\lambda(M) n$ fixed.}.
Independently of the detailed form of the field configuration furnishing the steepest descent, the right hand side of eq.~(\ref{eq:corr_lambda2}) will then take the form
\begin{equation}\label{eq:sqrtLambda}
\lambda_0^{-1/2}\,e^{\f{1}{\lambda_0}\Gamma_{-1}(\lambda_0 n, x_{fi}) +\Gamma_0(\lambda_0 n, x_{fi})+\lambda_0 \Gamma_1(\lambda_0 n, x_{fi})+\dots}\, ,\qquad x_{fi} =x_f - x_i.
\end{equation}
The factor $\lambda_0^{-1/2}$ is understood as follows. The path integral in the denominator is computed through a saddle-point expansion around the trivial point $\phi=\bar{\phi}=0$, while the action of the path integral in the numerator is stationary on a continuous family of nontrivial configurations with $\phi,\bar \phi\not = 0$ and parametrized by  the zero mode associated to the corresponding spontaneous breaking of the $U(1)$ symmetry. As the integral over the zero mode is clearly independent of the value of the action, this results in a mismatch of the powers of $\lambda_0^{1/2}$ in between the numerator and the denominator, leading to \eqref{eq:sqrtLambda} \footnote{The situation is fully analogous to the following example involving two dimensional integrals:
\begin{equation}\nn
I(\lambda,n)=\frac{\int_{\mathds{C}} dz d\bar{z}(z\bar{z})^{n} \exp\left\{-\frac{1}{\lambda}\left[z\bar{z}+\frac{1}{4}(z\bar z)^2\right]\right\}}{\int_{\mathds{C}} dz d\bar{z} \exp\left\{-\frac{1}{\lambda}\left[z\bar{z}+\frac{1}{4}(z\bar z)^2\right]\right\}}=\frac{\int_{\mathds{C}} dz d\bar{z} \exp\left\{-\frac{1}{\lambda}\left[z\bar{z}+\frac{1}{4}(z\bar z)^2-\lambda n\log (z\bar z )\right]\right\}}{\int_{\mathds{C}} dz d\bar{z} \exp\left\{-\frac{1}{\lambda}\left[z\bar{z}+\frac{1}{4}(z\bar z)^2\right]\right\}} .
\end{equation} 
The integral in the denominator is performed in an expansion around $z=\bar{z}=0$ and is thus proportional to $\lambda$ due to the gaussian integration on the two directions of the plane. The exponent in the numerator is instead stationary on the whole circle defined by $z\bar{z}=\sqrt{1+2\lambda n}-1$; in this case, while the integral over the radial direction produces a factor of $\sqrt{\lambda}$, angular integration gives an overall factor of $2\pi$. The full result, for arbitrary $\lambda n$, is thus proportional to $\lambda^{-1/2}$:
\begin{equation}\nn
I(\lambda,n)=\sqrt{\frac{2 \pi }{\lambda}} \frac{e^{-\frac{\lambda n+\sqrt{1+2 \lambda n}-1}{2 \lambda }} \left(\sqrt{1+2 \lambda n}-1\right)^{n+\frac{1}{2}}}{\l(1+2 \lambda n\r)^{1/4}}\left[1+\mO\l(\lambda\r)\right].
\end{equation}
}.

Now, notice that by using Stirling's formula the expression  $\lambda_0^{n+1/2} n!$ can be written in the same form as the exponential factor in eq. \eqref{eq:sqrtLambda}. It is then convenient to redefine the $\Gamma_k$'s  so as to factor out  a  $\lambda_0^{n+1/2} n!$ in the exponential factor in eq. \eqref{eq:sqrtLambda} and rewrite that equation as 
\be\label{gammaseries}
Z_{\phi^n}^2 \lambda_0^{n}\la [\bar{\phi}^n](x_f) [\phi^n] (x_i) \ra=
\lambda_0^{n} n! \,e^{\f{1}{\lambda_0}\Gamma_{-1}(\lambda_0 n, x_{fi}) +\Gamma_0(\lambda_0 n, x_{fi})+\lambda_0 \Gamma_1(\lambda_0 n, x_{fi})+\dots}
\ee
Comparing to eq.~(\ref{eq:corr_lambda}), we deduce that the exponential factor in eq.~(\ref{gammaseries}) 
coincides at weak coupling and finite $n$ with the loop expansion we discussed in the previous section.   In particular, given 
\be
D(x) = \f {1} {{ \Omega_{d-1} (d-2) (x^2)^{d/2-1}}}=\langle\bar{\phi}(x)\phi(0)\rangle_{free}\,,\qquad \Omega_{d-1}=\f{2 \pi^{d/2}}{\G(d/2)}
\ee
one has
\be
\lim _{\lambda_0\to 0}  e^{\f{1}{\lambda_0}\Gamma_{-1}(\lambda_0 n, x_{fi}) +\Gamma_0(\lambda_0 n, x_{fi})+\lambda_0 \Gamma_1(\lambda_0 n, x_{fi})+\dots} = D(x_{fi})^n\,.
\ee
Moreover one has that   the  $\lambda_0^\kappa\Gamma_\kappa$'s must possess a power series expansion in $\lambda_0$ with fixed $n$. Renormalization is simply performed by separating out the UV divergent part in each term in the exponent 
\be
\lambda_0^\kappa \G_\kappa (\lambda_0n,x_{fi})=\lambda^\kappa\G^{div}_\kappa(\lambda n,\lambda) +
\lambda^\kappa\G^{ren}_\kappa(\lambda n,\lambda,x_{fi}, M) 
\ee
where of course $\lambda\equiv \lambda(M)$ and where the resulting $\lambda^\kappa\bar \G_\kappa$ behave like power series at $\lambda=0$.
From eqs.~(\ref{eq:corr_lambda2},\ref{gammaseries}) we can then write
\be
\label{expZ}
Z_{\phi^n}^2=e^{\sum_{\kappa=-1}\lambda^\kappa\G^{div}_\kappa(\lambda n,\lambda)}\equiv e^{\sum_{\kappa=-1}\lambda^\kappa\bar\G^{div}_\kappa(\lambda n)}
\ee
and 
\be
\label{ren2point}
\la [\bar{\phi}^n](x_f) [\phi^n] (x_i) \ra=n! \,e^{\sum_{\kappa=-1}\lambda^\kappa\G^{ren}_\kappa(\lambda n,\lambda,x_{fi}, M)}\equiv n! \,e^{\sum_{\kappa=-1}\lambda^\kappa\bar\G^{ren}_\kappa(\lambda n,x_{fi}, M)}\,.
\ee
where, in the rightmost expressions, we rearranged the expansion in $\lambda$ using the (asymptotic) power series expansion of the $\lambda^\kappa \G_\kappa$. Eq.~(\ref{expZ}) provides a formal proof of eqs.~(\ref{gammastructure},\ref{gammastructure2}).
In the above expression the $\bar \G_\kappa$ represents the $(\kappa+1)$-loop correction to the saddle point approximation. In particular $\bar\Gamma^{div}_{-1}(\lambda n)$ and $\bar \Gamma^{ren}_{-1}$, represent  the  leading semiclassical contribution, the exponent at the saddle point\footnote{As we shall illustrate in a moment and, as it must be according to our derivation, the divergent part  appears from purely classical properties of the saddle point solution.}. However, they fully determine
the leading-$n$ contribution $F_0(\lambda n)$ in eq.~(\ref{gammastructure2}), thus resumming at once the largest powers of $n$ up to arbitrarily high-loop orders in the standard diagrammatic approach!
The remarkable result highlighted by our formal derivation and by eq.~(\ref{gammastructure2}), is that 
the result is organized as a 't Hooft expansion in which $\lambda n$ is the fixed 't Hooft coupling while $\lambda\ll 1$ and $n\gg 1 $. 

The rest of the paper is devoted to explicitly deriving these expressions, at leading (LO) and next-to-leading (NLO) order in the $\lambda$ expansion with $\lambda n$ fixed. In the next subsection we will perform a warm up computation by working at small but fixed $\lambda n$. In the later sections we shall develop the case of arbitrary $\lambda n$ by focussing on the Wilson-Fisher fixed point, where conformal invariance permits to tackle some technical difficulties in the computation.

\subsection{Semiclassics at small fixed \texorpdfstring{$\lambda n$}{lambda n} }

At small $\lambda n$ ordinary perturbation theory works. In this case the path integral eq.~(\ref{eq:corr_lambda})
can be computed by expanding around the trivial background $\phi=\bar\phi=0$. In that case the insertions of $\phi^n$ and $\bar\phi^n$, are not included in the exponent (as the exponent of eq.~(\ref{eq:corr_lambda2}) is singular at $\phi=\bar\phi=0$) and are purely determined by the quantum fluctuation $\delta \phi$  around the trivial solution, i.e. $\phi\equiv 0+ \delta \phi$. The loop expansion is purely generated by the small quartic term $\lambda\phi^4$. For instance, working at order $\lambda$ one finds 
%
%
\begin{equation}\label{eq:corrOneLoop}
\langle \bar \phi^n (x_f) \phi^n (x_i) \ra =  
\frac{n!\left[1- \frac{\lambda n(n-1)}{2(4\pi)^2} 
\left( \f{2}{\eps} +\log x_{fi}^2+ 1+\g + \log \pi  \right)+\mO\left(\frac{\lambda^2}{(4\pi)^4}\right)\right] }{\l[ \Omega_{d-1} (d-2) \r ]^n (x_{fi}^2)^{n \l(\f{d}{2}-1\r) }} .
\end{equation}
compatibly with the one-loop contribution to $\gamma_{\phi^n}$ derived in section 2. 

%

 As $\lambda n$ grows, the fluctuations of $\bar\phi^n (x_f) \phi^n (x_i)$ become significant, and for sufficiently large  $\lambda n$ they cannot be captured by perturbation theory. However eq.~(\ref{eq:corr_lambda2}) invites us to perform the computation around the stationary points of 
 \be\label{eq:FF_action_eff}
 S_{eff}\equiv \int d^d x \l [ \p \bar \phi \p \phi +\f{1} {4} \l ( \bar \phi \phi\r )^2\r ]-n\lambda_0\left (\log \bar \phi(x_f)+ \log \phi(x_i)\right).
 \ee
%
The equations of motion defining the stationary configuration include the operator insertions as a source
\bea
\p^2 \phi (x) -\f {1}{2} \phi^2(x)\bar \phi(x)& = & - \f {\lambda_0n} {\bar \phi (x_f)} \d ^ {(d)} (x-x_f), \nn \\
\p^2 \bar \phi (x) - \f {1}{2} \phi(x)\bar \phi^2(x)& = & - \f {\lambda_0n} {\phi (x_i)} \d ^ {(d)} (x-x_i).
\label{eq:saddleEq_lambda}
\eea 

Before discussing the details of the general computation, it is instructive to discuss the solution of \eqref{eq:saddleEq_lambda} for small $\lambda n$. Namely, we compute the function $\G_{-1}(\lambda n)$ in \eqref{gammaseries} to order $\mO\left(\lambda^2 n^2/(4\pi)^4\right)$
and we check that the result agrees with \eqref{eq:corrOneLoop}. As we work at first order in the coupling, in what follows we will take  $\lambda_0 = \lambda$. Now, for small $\lambda n$  the equations \eqref{eq:saddleEq_lambda} can be solved perturbatively; to this aim, it is convenient to expand the fields as
\begin{equation}
\phi=(\lambda n)^{1/2}\left[\phi^{(0)}+\phi^{(1)}+\ldots\right],\qquad
\bar{\phi}=(\lambda n)^{1/2}\left[\bar{\phi}^{(0)}+\bar{\phi}^{(1)}+\ldots\right],
\end{equation}
where $\phi^{(k)},\bar{\phi}^{(k)}=\mO\left(\lambda^k n^k\right)$. At the zeroth order, the equations of motion read
\bea
\p^2 \phi^{(0)}(x) & = & - \f {1} {\bar \phi^{(0)}(x_f)} \d ^ {(d)} (x-x_f), \nn \\
\p^2 \bar \phi^{(0)} (x) & = & - \f {1} {\phi^{(0)} (x_i)} \d ^ {(d)} (x-x_i), \label{eq:SaddleFree}
\eea
whose solution is uniquely defined up to one free parameter
and has the form
\bea
\phi^{(0)}(x) & = & \frac{c_0}{\Omega_{d-1} (d-2)} \f {1} {|x-x_f|^{d-2}}, \nn \\
\bar{\phi}^{(0)} (x) & = & \f {\bar{c}_0}{\Omega_{d-1} (d-2)} \f {1} {|x-x_i|^{d-2}};
\label{eq:FF_sol}
\eea
with the parameters $c_0$ and $\bar{c}_0$  related by 
\begin{equation}
c_0\bar{c}_0=\Omega_{d-1} (d-2)|x_f-x_i|^{d-2}.
\end{equation}
Notice that on the saddle-point, i.e. on the solution of \eqref{eq:SaddleFree}, the fields $\phi$ and $\bar{\phi}$ are analytically continued away from the original integration contour, since they are not related by complex conjugation. As a consequence, the fields appearing in the source terms in the right hand side of \eqref{eq:SaddleFree} have a finite value and no regularization procedure is needed to find the solution \eqref{eq:FF_sol}. Finally, the arbitrariness in the solution is related to the symmetry $(\phi,\bar{\phi})\rightarrow (\alpha\phi,\alpha^{-1}\bar{\phi})$ of the action \eqref{eq:FF_action_eff} analytically continued to arbitrary values of the fields. The one free parameter in the solution precisely corresponds to the presence of the one zero mode we mentioned before.

The next to leading contribution is determined by 
\bea
\p^2 \phi^{(1)} (x) & = & \f {\lambda n}{2} \left[\phi^{(0)}(x)\right]^2
\bar{\phi}^{(0)}(x)
+\frac{\bar{\phi}^{(1)}(x_f)}{\left[\bar{\phi}^{(0)}(x_f)\right]^2}
\delta^{(d)}(x-x_f), \nn \\
\p^2 \bar{\phi}^{(1)} (x) & = & \f {\lambda n}{2} \left[\bar{\phi}^{(0)}(x)\right]^2
\phi^{(0)}(x)
+\frac{\phi^{(1)}(x_i)}{\left[\phi^{(0)}(x_i)\right]^2}
\delta^{(d)}(x-x_i).
\eea
The solution reads
\bea
\phi^{(1)} (x) & = & -\f {\lambda n}{2} \int d^d y D(x-y)\left[\phi^{(0)}(y)\right]^2
\bar{\phi}^{(0)}(y)
-D(x-x_f)\frac{\bar{\phi}^{(1)}(x_f)}{\left[\bar{\phi}^{(0)}(x_f)\right]^2}, 
\nn \\
 \bar{\phi}^{(1)} (x) & = & -\f {\lambda n}{2} \int d^d y D(x-y)\left[\bar{\phi}^{(0)}(y)\right]^2
\phi^{(0)}(y)
-D(x-x_i)\frac{\phi^{(1)}(x_i)}{\left[\phi^{(0)}(x_i)\right]^2}, \label{eq:solFlatNLO}
\eea
where $\phi^{(1)}(x_i)$ and $\bar{\phi}^{(1)}(x_f)$ satisfy
\begin{equation}
\frac{\phi^{(1)}(x_i)}{c_0}+\frac{\bar{\phi}^{(1)}(x_f)}{\bar{c}_0}=-
\frac{\lambda n}{2}\int d^dyD^2(x_i-y)D^2(x_f-y).
\end{equation}
There is a one parameter arbitrariness in the solution due to the aforementioned symmetry. The integrals are formally divergent in $d=4$ and thus are performed via standard dimensional regularization techniques. Plugging the solution in the action \eqref{eq:FF_action_eff}, we find
\bea
S_{eff} & = & \lambda n - \lambda n \log \l [\f {\lambda n}{\Omega_{d-1} (d-2)} \f {1} {(x_{fi}^2)^{d/2-1}} \r ] \nn \\
& + & 
\lambda^2 n^2 \l ( \f{1}{16 \pi^2 \varepsilon} + \f {1+\g + \log \pi }{32\pi^2} \r ) + \f{\lambda^2 n^2}{32\pi^2} \log x_{fi}^2\,.
\eea
$e^{-S_{eff}/\lambda}$  must represent   the leading term 
\be
\lambda^{n}n! e^{\frac{\G_{-1}}{\lambda}}
\ee
 in eq.~(\ref{gammaseries}) with $\G_{-1}$ expanded up to $\mO(\lambda^2n^2)$. It is easy to see it does. In particular, $\log n! \approx n \log n - n$ ensures that $\G_{-1}$ has a well defined power series in $\lambda n$ as expected. The correlator, according to eqs.~(\ref{eq:corr_lambda},\ref{eq:corr_lambda2}), then reads
\be
\la \bar \phi^n (x_f) \phi^n (x_i) \ra = \f{n^n e^{-n} \exp\l [- \lambda n^2 \l( \f{1}{16 \pi^2 \eps} + \f {1+\g + \log \pi }{32\pi^2} \r ) \r]} {\l [ \Omega_{d-1} (d-2) \r ]^n (x_{fi}^2)^{n \l(\f{d}{2}-1\r)+\f{\lambda n^2}{32\pi^2} }}.
\label{eq:pertPlaneCorr}
\ee
This expression\footnote{This expression was recently derived also in \cite{Arias-Tamargo:2019xld}, where the authors considered the correlator in the $\lambda\rightarrow 0$ limit with $\lambda n^2$ fixed, clearly corresponding to small $\lambda n$. This is just a particular limit of the general formula \eqref{ren2point}, as our approach makes clear.} reproduces the result of the standard perturbative computation \eqref{eq:corrOneLoop} up to subleading terms at large $n$. 
Remarkably, the $\mO(\lambda n^2)$ correction to the scaling dimension results in \eqref{eq:corrOneLoop}  from a genuine one-loop computation, while it results in \eqref{eq:pertPlaneCorr}  from the \emph{classical} solution of the saddle point equations \eqref{eq:saddleEq_lambda}. According to our discussion, the subleading $\mO(\lambda n)$ contribution to $\gamma_{\phi^n}$ in eq.~\eqref{eq:corrOneLoop}, would instead arise from the first quantum correction around the saddle, i.e. from $\G_0$ in eq.~(\ref{gammaseries}). Our alternative semiclassical computation shows that the $\mO(\lambda n^2)$ contribution to $\gamma_{\phi^n}$ is a genuinely classical contribution, while the $\mO(\lambda n)$ is intrinsically quantum. The emergence of classical physics in the presence of large quantum numbers, $n$ in this case, is a crucial fact of physics. Our case here is closely analogous to the relation between the classical approximation to the squared angular momentum, $\ell^2$, and  the exact quantum result, $\ell(\ell+1)$ (see ref.~\cite{Monin:2016jmo} for an illustration).



\section{Finite \texorpdfstring{$\lambda n$}{lambda n} on the cylinder}\label{secCylinder}
Finding the general solution of \eqref{eq:saddleEq_lambda} is in general a technically challenging task, but symmetries can help tackle the difficulties. In the case at hand the relevant ones are $U(1)$ symmetry, rotational invariance and dilations. Starting with $U(1)$, the conservation of the associated Noether current
%
%
%
\be\label{eq:current}
j_\m = \bar \phi \p _ \m \phi - \phi \p_\m \bar \phi.
\ee
provides powerful insight.
The field insertions in \eqref{eq:FF_action_eff} act as a source for the current \eqref{eq:current}. Indeed, from the equations of motion \eqref{eq:saddleEq_lambda} we get
\be
\p _ \m j^\m = n \d^{(d)} (x-x_i) - n \d^{(d)} (x-x_f).
\ee
We can then use Gauss law to determine the flux of the current through a sphere centered at $x_i$ with radius $r$:
\be
\oint_{x_i}  \, d \Omega_{d-1} \,r^{d-1} j^\m(x) n_\m(x) = n\,
\theta\left(|x_f-x_i|-r\right),
\label{eq:chargePlaneConstr}
\ee
where $n_\m(x)$ is the unit vector orthogonal to the sphere at point $x$. Sufficiently close to the point $x_i$, i.e. for $| x-x_i | \ll | x_f-x_i |$, we expect the solution of \eqref{eq:saddleEq_lambda} to be approximately spherically symmetric. In this regime, we then conclude from eq. \eqref{eq:chargePlaneConstr} that the current is given by
\be
j_\m(x) = \f {n} {\Omega_{d-1}} \f{(x-x_i)_\m}{|x-x_i |^d} \l [ 1 + \mO\l(\f{|x-x_i |}{|x_f-x_i |} \r) \r ].
\label{eq:currApprox}
\ee
This equation provides a simple constraint involving both $\phi$ and $\bar \phi$. Unfortunately it is not enough to fix their coordinate dependence. In fact, even in the regime $| x-x_i | \ll | x_f-x_i |$, where spherical symmetry is expected, the radial dependence of the solution is non-trivial, as one can convince oneself by making eq.~(\ref{eq:solFlatNLO}) explicit. The origin of such a complicated dependence is the lack of dilation invariance
of generic $\lambda \phi^4$ in $d$-dimension. Notice, instead, that in the free case, where dilations are a symmetry, the solution displays a simple scaling behaviour. Working in strictly $d=4$, where $\phi^4$ is scale invariant is also not an option,  because of the need for regulation\footnote{If we contented ourselves with the leading semiclassical approximation we could work in $d=4$ and regulate $\phi^n$ by point splitting.}. We thus conclude that the only way forward in order to more easily derive the solution is to work directly at the Wilson-Fisher fixed point, where we can profit from the bonus of scale invariance. That also matches well, and not unrelatedly,  the fact that only at the fixed point is the anomalous dimension  a fully physical quantity.

%
%

\subsection{Weyl map to the cylinder}

The advantage of working at the fixed point is that we can exploit the power of conformal invariance.  That allows to map our theory from the plane to the cylinder 
\be
\mathbb{R}^d \to \mathbb{R} \times S^{d-1}\,, \label{eq:Weyl}
\ee
in such a way that the dilations on the plane are mapped  to  time translations on the cylinder.
Correspondingly, the spectrum of operator dimensions on the plane, the eigenvalues of the dilation charge $D$, are mapped
to the energy spectrum on the cylinder, the eigenvalues of $H_{cyl}$. Our goal of computing the dimension of $[\phi^n]$ is thus mapped into the computation of the energy of the corresponding state on the cylinder. The advantage offered by this viewpoint is that time translations on the cylinder, unlike dilations on the plane, are a symmetry also away from the fixed point. When mapping  our  semiclassical  computation to the cylinder, we will thus have an additional symmetry controlling the classical solution, even away from criticality. In other words, while, in the approach of the previous section, a simple scaling ansatz for the radial dependence of the solution is  inconsistent, given the lack of scale invariance in the regulated theory, on the cylinder it is possible to consistently look for a solution that is stationary in time. That enormously simplifies our task. Of course, we must stress that this very non trivial simplification only works at the fixed point.

In what follows we briefly review the mapping of our computation to the cylinder. A more detailed discussion of this subject, including the operator state correspondence, can for instance be found in~\cite{Rychkov:2016iqz,Simmons-Duffin:2016gjk}.

Parametrizing $\mathbb{R}^d$ by polar coordinates $(r,\Omega_{d-1})$, where  $\Omega_{d-1}$  collectively denotes the coordinates on $S^{d-1}$, and $\mathbb{R} \times S^{d-1}$ by $(\tau,\Omega_{d-1})$, the mapping is simply given by $r=Re^{\tau/R}$ with $R$ the sphere radius. The cylinder metric is then related to the flat one by a Weyl rescaling
\begin{equation}
ds^2_{cyl}=d\tau^2+R^2d\Omega^2_{d-1}=\frac{R^2}{r^2}ds^2_{flat}\,.
\end{equation}
 The action of the theory on the cylinder reads\footnote{From this point forward we will be working with canonically normalized fields.}
\begin{equation}
S_{cyl}= \int d^dx\sqrt{g} \l [ g^{\mu\nu} \p_\m \bar \phi \p _\n \phi + m^2 
 \bar \phi  \phi +
\f{\lambda_0}{4} \l ( \bar \phi  \phi\r )^2\r ], \label{eq:cylAction}
\end{equation}
where the mass term $m^2=\l(\frac{d-2}{2R}\r)^2$ arises from the ${\cal R}(g)\bar \phi\phi$ coupling to the Ricci scalar which is enforced by conformal invariance\footnote{Hence, at the fixed point, $m^2$ is not renormalized by loop effects.}~\cite{Brown:1980qq}. 

Weyl invariance\footnote{The Weyl anomaly does not affect correlation functions of local operators \cite{Simmons-Duffin:2016gjk}.} at the fixed point ensures that the flat space theory \eqref{eq:phi4Lagrangian} is equivalent to the one on the cylinder described by \eqref{eq:cylAction}. In particular, the two-point function of a scalar primary operator $\mO$ of scaling dimension $\Delta_{\mathcal{O}}$ and its conjugate on the cylinder is related to the flat space one by \cite{Simmons-Duffin:2016gjk,Rychkov:2016iqz}
\begin{equation}\label{eq:GenericCorrelatorCyl}
\langle\mathcal{O}^{\dagger}(x_f)\mathcal{O}(x_i)\rangle_{\text{cyl}}=
|x_f|^{\Delta_{\mathcal{O}}}|x_i|^{\Delta_{\mathcal{O}}}
\langle\mathcal{O}^{\dagger}(x_f)\mathcal{O}(x_i)\rangle_{\text{flat}}\equiv \frac{|x_f|^{\Delta_{\mathcal{O}}}|x_i|^{\Delta_{\mathcal{O}}}}{ |x_f-x_i|^{2\Delta_{\mathcal O}}}.
\end{equation}
where in the last two equations $x_{f,i}$ are meant to represent the Cartesian coordinates on the plane. We also assumed ${\mathcal O}$ to be canonically normalized.
 Now, the limit $x_i \to 0$ on the plane translates to $\tau_i\to -\infty$ on the cylinder and the above equation becomes
\be\label{xiinfinity}
\langle\mathcal{O}^{\dagger}(x_f)\mathcal{O}(x_i)\rangle_{\text{cyl}}\overset{\tau_i\rightarrow-\infty}{=} 
e^{-E_{\mathcal O}(\tau_f- \tau_i)},\qquad
E_{\mathcal O}=\Delta_{\mathcal O}/R\,.
\ee
More precisely one can check that the rate of approach to the above limiting result is controlled by $e^{\tau_i/ R}$. So that the above equation holds with exponential precision for $|\tau _i/ R|\gg 1$.
By eq.~(\ref{xiinfinity}) the action of  $\mathcal{O}(x_i)$ at $\tau_i\to -\infty$ simply creates a state with energy $\Delta_{\mathcal O}/R$ and carrying all the  global quantum numbers of ${\mathcal O}$. This is the 
operator state correspondence, which greatly illuminates many aspects of conformal field theory when viewed on the cylinder.

In the following, we shall consider $\mathcal{O}=\phi^n$, and $\mathcal{O}^\dagger = \bar \phi^n$,  work only on $\mathbb{R}\times S^{d-1}$ and hence  drop the subscript cyl.
By the same argument as just above, the two-point function $\la \bar{\phi}^n (x_f) \phi^n (x_i) \ra$, with  $\tau _ {f,i} = \pm T /2$, for $T\rightarrow\infty$ directly yields the scaling dimension $\Delta_{\phi^n}$ 
\be
\la \bar{\phi}^n (x_f) \phi^n (x_i) \ra
\overset{T\rightarrow\infty}{=} \mc N e^{-E_{\phi^n} T},\qquad
E_{\phi^n}=\Delta_{\phi^n} /R,
\label{eq:groundState}
\ee
where the (divergent) coefficient $\mc N$ is independent of $T$. 

To compute the two point function we can then proceed with the methodology discussed at the beginning of section 3. The result  will have the structure of eq.~(\ref{gammaseries}). 
 Upon separating out the divergent and finite part of the $\lambda_0^\kappa \G_\kappa$'s, we will have a $T$ independent divergent piece determining the normalization factor $\mc N$, while the $T$ dependent part will be  finite when written in terms of $\lambda(M)$ and linear in $T$ for $T\gg R$. The linearity in $T$ will follow provided the solution is stationary in time, which it will be, thanks to time translation invariance of the action regardless of the theory being at the fixed point.
 Similarly to eq.~(\ref{ren2point}) we shall thus have
\bea \nn
RE_{\phi^n}&=&\frac{1}{\lambda_0}e_{-1}(\lambda_0 n,d)+e_{0}(\lambda_0 n,d)+\lambda_0 e_{1}(\lambda_0n,d)+\ldots\\
&=&\frac{1}{\lambda}\bar{e}_{-1}(\lambda n,RM,d)+\bar{e}_{0}(\lambda n,RM,d)+\lambda \bar{e}_{1}(\lambda n,RM,d)+\ldots,
\label{eq:cylinderEnergyExpansion}
\eea
where $\lambda\equiv\lambda(M)$ and $\bar{e}_k$ is defined from the $e_k$'s analogously to $\bar{\Gamma}_k$ in eq. \eqref{expZ}. By choosing $\lambda=\lambda_*$ the dependence on $RM$ will have to drop by scale invariance giving a result of the form
\be
\Delta_{\phi^n}=\frac{1}{\lambda_*}\Delta_{-1}(\lambda_* n)+\Delta_{0}(\lambda_* n)+\lambda_* \Delta_{1}(\lambda_* n)+\ldots\,.
\label{eq:cylinderDeltaExpansion}
\ee
In the remaining sections of the paper we shall explicitly compute the leading semiclassical contribution $\Delta_{-1}$ and the  first quantum correction $\Delta_{0}$.

\subsection{Leading order: \texorpdfstring{$\Delta_{-1}$}{Deltaminus1}}

In this section we compute the dimension $\Delta_{\phi^n}$ at the leading order in $\lambda$ using the operator state correspondence described above. More precisely, at this order, we shall compute the dimension of the lowest dimension operator with charge $n$ as a function of $\lambda n$. For sufficiently small $\lambda n$, such operator of lowest dimension  obviously coincides with $\phi^n$ as  shown by a perturbative analysis. Indeed, any other operator with charge $n$, such as
$\phi^{n-2}(x)  (\p \phi(x))^2$,
clearly possesses a larger scaling dimension in the free limit, and for small enough $\lambda n$ the ordering is not affected. Level crossing may in principle occur at finite $\lambda n$, but that would unavoidably be associated with a non-analyticity in the dependence on $\lambda_*n$ of the minimal dimension at fixed charge.
The result we shall obtain with our semiclassical method is however analytic at positive $\lambda_* n$ and matches the dimension of $\phi^n$ at small $\lambda_*n$. Thus we conclude our result represents the dimension of $\phi^n$ at arbitrary $\lambda_* n$, justifying \emph{a posteriori} our approach.

%
%
%

Having said that, we further proceed along the lines of~\cite{Monin:2016jmo}. Namely, we compute the expectation of the 
evolution operator $e^{-HT}$ in an arbitrary state $| \psi _n  \ra$ with fixed charge $n$. As long as there is an overlap between the state $| \psi _n \ra$
and the lowest energy state (with charge $n$), in the limit $T \to \infty$ the expectation gets saturated by the latter 
\be
\la \psi_n | e^{-HT} | \psi _n \ra \underset{T\to \infty}{=} \tilde{ \mc N} e^{-E_{\phi^n} T}.
\label{eq:Evolution}
\ee
Now the choice of the state $| \psi _n  \ra$ is completely in our hands and we take it to be
\be
| \psi _n  \ra = \int \mc D \a (\vec n) \, \exp \l [ i \frac{n}{R^{d-1}\Omega_{d-1}} \int d\Omega_{d-1} \, \a (\vec n) \r ] |f,\a(\vec n) \ra,
\ee
where $\vec{n}$ denotes collectively the coordinates on the $d-1$ dimensional sphere and the state $|f,\a(\vec n) \ra$ is the one with fixed values of the fields\footnote{The fields are independent of $\tau$ as the state is defined in Schr\"{o}dinger picture.} $\rho(\vec n)=f$ and $\chi(\vec n)=\a(\vec n)$ defined as
\be
\phi = \f{\rho}{\sqrt{2}} e^{i \c}, ~~ \bar\phi = \f{\rho}{\sqrt{2}} e^{-i \c}.
\label{eq:waveFunction}
\ee
The result for $E_{\phi^n}$ is independent of the constant value $f$, however, a specific choice, that will be derived later, makes computations much simpler.
Plugging (\ref{eq:waveFunction}) into (\ref{eq:Evolution}) and using the path integral representation for the evolution operator we obtain
\be\label{eq:PI1}
\la \psi_n | e^{-HT} | \psi _n \ra = \mathcal{Z}^{-1}\int \mc D \c_i \mc D \c_f e^ {-i \frac{n}{R^{d-1}\Omega_{d-1}} \l [ \int d\Omega_{d-1} (\c _f - \c_i )\r ] } \int^{\rho=f,~ \c=\c_f}_{\rho=f, ~\c=\c_i} \mc D \rho \mc D \c e^{-S},
\ee
where we defined
\begin{equation}\label{eq:PInormalization}
\mathcal{Z}=\int \mc D \phi \mc D \bar \phi \, e^{-S} ,
\end{equation}
ensuring that the vacuum to vacuum amplitude is normalized to unity, 
$\langle 0|E^{-HT}|0\rangle=1$.
Using that the boundary conditions imply
\be
\int d\Omega_{d-1} (\c _f - \c_i ) = \int^{T/2}_{-T/2} d\tau \int d\Omega_{d-1} \dot \c
\ee
where $\dot{\chi}\equiv\p_\tau\chi$, eq. \eqref{eq:PI1} can be rewritten as a finite time path integral with boundary conditions only for $\rho$ :
\be\label{eq:PI2}
\la \psi_n | e^{-HT} | \psi _n \ra =\mathcal{Z}^{-1}\int^{\rho=f}_{\rho=f} \mc D \rho \mc D \c e^{-S_{eff}},
\ee
where the action on the right hand side is given by
\be
S_{eff} =  \int_{-T/2}^{T/2} d\tau\int d\Omega_{d-1} \l [ \f{1}{2} (\p \rho)^2 +  \f{1}{2} \rho^2 ( \p \c ) ^ 2 + \f{m^2}{2} \rho^2 +
\f{\lambda_0}{16} \rho^4 +i\frac{n}{R^{d-1}\Omega_{d-1}}\,\dot{\chi}\r ]\,.
\label{eq:effActCyl}
\ee

We can now perform the path integral in \eqref{eq:PI2} via a saddle point approximation. The variation of the action \eqref{eq:effActCyl} provides the equations of motion for the fields
\begin{equation}\label{eq:saddleChi}
-\p^2\rho+\l [ (\p\c)^2 +m^2 \r ] \rho+ \f{\lambda_0}{4}\rho^3=0,\qquad
i\p_\m \l ( \rho^2  g^{\m\nu}\p _\n\c \r )  =0,
\end{equation}
supplemented by the following condition which fixes the value of the charge
\begin{equation}\label{eq:fixCharge}
i\rho^2\dot{\chi}=\frac{n}{R^{d-1} \Omega_{d-1}}.
\end{equation}
By a proper choice of the initial and final value $\rho_i=\rho_f=f$ in the wave-function, the stationary configuration for the action \eqref{eq:effActCyl} takes the following simple form
\be
\rho = f\,, \qquad\qquad \c= -i \m \tau +\text{const.}\,,
\label{eq:cylinderSol}
\ee
where the constants $f$ and $\m$ are fixed by the first equation in \eqref{eq:saddleChi} and by \eqref{eq:fixCharge}
\be
(\m^2-m^2)=\f{\lambda_0 }{4}f^2, \qquad\qquad
\m f^2 R^{d-1} \Omega_{d-1} = n.
\label{eq:mu_f_equations}
\ee
Given the constraint $f^2\geq 0$, imposed by the boundary condition $\rho_i=\rho_f=f\in {\mathbb R}$,
these equations admit a unique solution for $f^2$ and $\mu$. On this profile $\chi$ is analytically continued to the complex plane (see the comments below \eqref{eq:FF_sol}). Notice that the condition $f^2\geq 0$ implies that the solution for $\mu$ is discontinuous at $\lambda_0 n=0$. This can be seen easily substituting $f^2\propto n/\mu$ in the first equation in \eqref{eq:mu_f_equations}:
\begin{equation}\label{eq:mu_f_equations_extra}
\mu(\mu^2-m^2)=\frac{\lambda_0n}{4R^{d-1}\Omega_{d-1}}\quad\text{with}\quad n/\mu\geq 0,
\end{equation}
where the last inequality follows from the reality condition on $f$. It is then obvious that the, otherwise analytical, solution of \eqref{eq:mu_f_equations_extra} satisfies
$\mu(\lambda_0 n)=-\mu(-\lambda_0 n)$, implying the existence of a discontinuity for $\lambda_0n\rightarrow 0$, where $\mu\simeq \text{sgn}(n)\left[m+\mO\left(\lambda_0 n\right)\right]$.
As a consequence of the latter, also the scaling dimension $\Delta_{\phi^n}$ will be non-analytic at $\lambda n=0$. This reflects the physical fact that the scaling dimension of $\phi^n$ and the operator with opposite charge, $\bar{\phi}^n$, are the same; as the expansion \eqref{gammastructure} contains odd powers of $n$, the physical scaling dimension cannot be continuous at $n=0$.
In the following, we implicitly consider only $n>0$.

Physically, the solution \eqref{eq:cylinderSol} describes a superfluid\footnote{This means that there is a linear combination of $U(1)$ transformations and time ($\tau$) translations which leaves invariant the configuration \eqref{eq:cylinderSol} \cite{Nicolis:2011cs}.} phase \cite{Son:2002zn}, with homogeneous charge density $j_0=\mu f^2$ and chemical potential given by $\mu$. The action \eqref{eq:effActCyl} evaluated on such configuration provides the leading order value for the energy \eqref{eq:cylinderEnergyExpansion}:
\begin{equation}\label{eq:Eminus1}
\f{1}{\lambda_0}\frac{e_{-1}(\lambda_0 n,d)}{R}=S_{eff}/T = \frac{n}{2}  \l ( \frac32\m +\frac12 \f {m^2}{\m}\r ).
\end{equation}
Had we chosen $\rho_i,\rho_f\neq f$, $\rho(\tau)$ would have approached exponentially fast the value $\rho=f$ away from the boundaries. As a result, in the $T\rightarrow\infty$ limit the contribution of the action growing linearly in time is independent of the precise value of the boundary conditions for $\rho$.

To obtain the leading order $\Delta_{-1}$ in \eqref{eq:cylinderDeltaExpansion}, we consider the classical value for the chemical potential obtained from \eqref{eq:mu_f_equations} setting $\lambda_0=\lambda_*$ and $d=4$ everywhere else:
\be
R\m_*  = \frac{3^{1/3}+\left[9 \frac{\lambda_* n}{(4\pi)^2}-\sqrt{81 
\frac{\left(\lambda_* n\right)^2}{(4\pi)^4}-3}\right]^{2/3}}{3^{2/3} \l[9 \frac{\lambda_* n}{(4\pi)^2}-
\sqrt{81 \frac{\left(\lambda_* n\right)^2}{(4\pi)^4}-3}\r]^{1/3} }.
\label{eq:mu}
\ee
Taking the complex conjugate of this expression, one can check that $R\m^*$ is real for $\lambda_* n\geq 0$.
Plugging in \eqref{eq:Eminus1} and taking $m=1/R$ we conclude that the classical contribution to the scaling dimension is
\begin{equation}
\frac{1}{\lambda_*}\Delta_{-1}=n F_{0}\l(\lambda_* n\r), \label{eq:DeltaTree}
\end{equation}
where the function $F_{0}$ reads:
\bea \nn
F_{0}(16\pi^2 x ) & =&
\frac{3 \l[9 x-\sqrt{81 x^2-3}\r]^{1/3}+3^{2/3}\l[9x- \sqrt{81 x^2-3}\r]}{\left[\left(9 x-\sqrt{81 x^2-3}\right)^{2/3}+3^{1/3}\right]^2}
\\ &+&
\frac{9 \times 3^{1/3} x\l[9x-  \sqrt{81x^2-3}\r]^{2/3}}{2\left[\left(9 x-\sqrt{81 x^2-3}\right)^{2/3}+3^{1/3}\right]^2}.
\label{eq:explicitFminus1}
\eea
Though not obvious, for $x>0$ this is a real and positive function, which grows monotonically with $x$. Remarkably, eq. \eqref{eq:DeltaTree}  explicitly resums the contribution of infinitely many Feynman diagrams.

The form of the result becomes particularly simple (and interesting) in the two extreme regimes, $\lambda_* n \ll (4\pi)^2$ and $\lambda_* n \gg (4\pi)^2$, where eq. \eqref{eq:DeltaTree} reads
\be
\frac{\D_{-1}}{\lambda_*} = 
\l \{
\ba{ccc} 
\dst n \l [ 1 + \f{1}{2} \, \l ( \f {\lambda_* n}{16\pi^2}\r ) - \f{1}{2} \,
 \l ( \f {\lambda_* n}{16\pi^2}\r ) ^2
  + \mO \l(\frac{(\lambda_* n)^3}{(4\pi)^6}\r) \r] , &\text{for}& \lambda_* n \ll (4\pi)^2, \\ \\
\dst \f{8\pi^{2}}{\lambda_*} \l [ \f{3}{4} \l( \f{\lambda_* n}{ 8\pi^{2}} \r ) ^{4/3} + \frac{1}{2}\l( \f{\lambda_* n}{ 8\pi^{2}} \r ) ^{2/3} + 
\mO \l(1\r) \r ], &\text{for}& \lambda_* n \gg (4\pi)^2.
\ea
\r.
\label{eq:DeltaMinus1}
\ee
The first line of \eqref{eq:DeltaMinus1} reproduces the result \eqref{eq:gammaTwoLoop} up to higher orders and thus provides a non trivial check of our approach. 
Notice that the agreement is independent of the precise value of $\lambda_*$, since at tree-level the Lagrangian \eqref{eq:phi4Lagrangian} is Weyl invariant for every value of the coupling and the theory can be safely mapped to the cylinder through a change of coordinates and a field redefinition.
In the opposite regime, the result is organized as an expansion in powers of $(\lambda_* n)^{2/3} $, in agreement with the predictions of the large charge expansion in CFT \cite{Hellerman:2015nra,Monin:2016jmo}.

The parameter which marks the difference between the two regimes is the chemical potential $\mu_*$, since, as we will see explicitly in the next section, the latter controls the gap of the radial mode. For small $\lambda_* n$ the chemical potential, is of order of $R^{-1}$, while in the opposite regime its value is proportional to $j_0^{1/3}\gg R^{-1}$. In this regime, the fact that the leading contribution in the second line of \eqref{eq:DeltaMinus1}  scales as $(\lambda_* n)^{4/3}/R\sim j_0^{4/3}R^3$ follows just from dimensional analysis \cite{Hellerman:2015nra}.

\subsection{One-loop correction: \texorpdfstring{$\Delta_0$}{Delta0} \label{sec:loopCorrCyl}}

Let us now compute the first subleading correction $\Delta_0$. To this aim we expand the fields around the saddle point configuration:
\be
\rho(x) = f +r(x), \qquad\c(x) = -i\mu \tau + \f{1}{f \sqrt{2}}\pi(x).
\ee
The action \eqref{eq:effActCyl} at quadratic order in the fluctuations reads
\be
S^{(2)} = \int^{T/2}_{-T/2} d\tau\int d\Omega_{d-1} \l [ \f{1}{2} (\p r)^2 + \f{1}{2} (\p \pi)^2 - 2 i\m \, r \p_\tau \pi + (\mu^2-m^2) r^2\r ].
\label{eq:cylAction2}
\ee
This action describes a gapped and a gapless mode, with dispersion relations given by
\be
\omega^2_\pm (\ell) = J^2_\ell+3\m^2-m^2\pm \sqrt{4 J^2_\ell\m^2+(3\m^2-m^2)^2},
\label{eq:dispersions}
\ee
where $J^2_\ell = \ell (\ell+d-2)/R^2$ is the eigenvalue of the Laplacian on the sphere. The gapless mode is the Goldstone boson for the spontaneously broken $U(1)$ symmetry. The gap of the first mode is:
\begin{equation}
\omega ^2_+(0)=6\mu^2-2m^2.
\end{equation}
Notice also that the $\ell=1$ excitation of the gapless mode has unit energy, $\omega_{-}(1)=1/R$ and corresponds to a descendant state.

As anticipated, in the large $\lambda n$ limit the gap of the radial mode grows as $(\lambda n)^{1/3}/R$. Henceforth, in this regime we can integrate out this mode and the lightest states at charge $n$ are described by an effective theory for the Goldstone mode only. The form of the effective theory was used in \cite{Hellerman:2015nra,Monin:2016jmo} to study the spectrum at large charge in a generic $U(1)$ invariant CFT and derive the form of the expansion in the second line of \eqref{eq:DeltaMinus1}. In this regime, the squared sound speed of the Goldstone mode, given by
\begin{equation}
\l(\frac{d\omega^2_-}{d J^2_\ell}\r)_{\ell=0}=\frac{\mu^2-m^2}{3\mu^2-m^2},
\end{equation}
approaches the value $1/3$ dictated by scale invariance in a fluid. 

To extract the first correction to the energy \eqref{eq:cylinderEnergyExpansion} we consider the one-loop expression for the path-integral \eqref{eq:PI2}:
\bea\nn
\langle\psi_n|e^{-HT}|\psi_n\rangle &=&
e^{-\frac{e_{-1}(\lambda_0 n,d) T}{\lambda_0 R} }
 \frac{\int \mc D r \mc D \pi \, \exp \l [ - S^{(2)} \r ]}{\int \mc D \phi \mc D \bar \phi \, \exp \l [ -\int^{T/2}_{-T/2} \l(\p \phi\p \bar \phi 
 +m^2\phi\bar{\phi}\r)\right] }\\
& = &\tilde{\mathcal{N}} \exp\l\{-\l[\frac{1}{\lambda_0}e_{-1}(\lambda_0 n,d)+e_0(\lambda_0 n,d)\r]\frac{T}{R}\r\},
 \label{eq:PI1LoopCyl}
\eea
where the normalization factor $\tilde {\mc N}$ is $T$-independent. The latter contains a factor $\lambda_0^{-1/2}$ coming form the integration over the zero mode (see the comments below \eqref{eq:sqrtLambda}). 
The denominator in the first line of \eqref{eq:PI1LoopCyl} arises from the normalization factor \eqref{eq:PInormalization}. In the second line, the correction to the energy arises from the fluctuation determinant of the Gaussian integrals in the numerator and the denominator. It can be written explicitly in terms of the expressions \eqref{eq:dispersions} and the formula for the free dispersion relation $\omega_0^2(\ell)=
J_{\ell}^2+m^2=\left(\ell+\frac{d-2}{2}\right)^2/R^2$: 
\begin{align}\nn
T\frac{e_0}{R}=\log \f{\sqrt{\det S^{(2)}}}{\det\l (- \p_\tau^2-\D_{S^{d-1}} +m^2\r )}
& = 
\frac{T}{2} \sum_{\ell=0}^{\infty} n_\ell\int \f{d\omega}{2\pi} \log \frac{\l[\omega^2+\omega_-^2(\ell)\r] 
\l[\omega^2+\omega_-^2(\ell)\r]}{\l [\omega^2+\omega_0^2(\ell) \r ]^2}  \\
 &= \frac{T}{2}\sum_{\ell=0}^\infty n_{\ell}\left[\omega_+(\ell)+\omega_-(\ell)-2\omega_0(\ell)\right], \label{eq:one-loop-det1}
\end{align}
where $n_\ell$ is the multiplicity of the Laplacian on the $(d-1)$-dimensional sphere:
\be
n_\ell = \f{(2\ell+d-2) \G(\ell+d-2)}{\G(\ell+1) \G(d-1)}.
\ee
In $d=4$ the multiplicity is $n_\ell=(1+\ell)^2$. In dimensional regularization, we can use the following identities which hold for sufficiently negative $d$
\begin{equation}
\sum_{\ell=0}^\infty n_{\ell}=\sum_{\ell=0}^\infty n_{\ell}\,\ell=0
\qquad\implies\qquad
\sum_{\ell=0}^\infty n_\ell \,\omega_0(\ell)=0.
\end{equation}
Finally we formally find the second term in the expansion \eqref{eq:cylinderEnergyExpansion} as a sum of zero point energies, as it could have been intuitively expected:
\begin{equation}\label{eq:E1loop}
e_0(\lambda_0n,d)=\frac{R}{2}\sum_{\ell=0}^{\infty}
 n_{\ell}\left[\omega_+(\ell)+\omega_-(\ell)\right].
\end{equation}

We can now compute the leading correction to the scaling dimension \eqref{eq:cylinderDeltaExpansion}. The details of the calculation are given in the appendix \ref{app:1loopCylsmall}. The result is formally written in terms of the classical value of the chemical potential \eqref{eq:mu} and a convergent infinite sum:
\begin{equation}\label{eq:Delta1loop}
\Delta_{0}= -\frac{15 \mu_*^4  R^4+6 \mu_*^2  R^2-5}{16}
+\frac{1}{2} \sum_{\ell=1}^\infty\sigma(\ell)
+\frac{\sqrt{3\mu_*^2R^2-1}}{\sqrt{2}},
\end{equation}
where $\sigma(\ell)$ is obtained by subtracting the divergent piece from the summand in \eqref{eq:E1loop}
\begin{equation}
\sigma(\ell)=(1+\ell)^2R\left[\omega^*_+(\ell)+\omega^*_-(\ell)\right]-
2\ell^3-6\ell^2-\l(2 \mu_*^2 R^2+4\r)\ell
-2R^2\mu_*^2+\frac{5 \left(\mu_*^2 R^2-1\right)^2}{4\ell}.
\end{equation}
As in equation \eqref{eq:mu}, the star stresses that all quantities are evaluated setting $\lambda_0=\lambda_*$ and $d=4$ everywhere else.

In the small $\lambda_* n$ limit, we can compute the sum in \eqref{eq:Delta1loop} analytically and we find
\begin{equation}
\Delta_{0}=-\frac{3 \lambda_*  n}{(4\pi)^2 }
+\frac{\lambda_*^2 n^2}{2(4\pi )^4 }
+\mO\l(\frac{\lambda_*^3 n^3}{(4\pi)^6}\r).
\end{equation}
Summing this to the leading order result \eqref{eq:DeltaMinus1} and recalling the relation between the coupling and the number of space dimensions \eqref{eq:lambda}, we determine $\Delta_{\phi^n}$ as:
\be
\D_{\phi^n} = n \l ( \f {d} {2} -1 \r )+\f{\varepsilon }{10} \,n (n-1)- \f{\varepsilon^2 }{50} n(n^2-4n)+\mO\l(\eps^2n,\eps^3n^4\r).
\ee
This is in perfect agreement with the diagrammatic calculation in eq. \eqref{eq:DeltaTwoLoop}. 

In the large $\lambda_* n$ limit the result \eqref{eq:Delta1loop} develops a contribution proportional to $\log(\lambda_* n)$, which arises from the divergent
tail of the sum in \eqref{eq:E1loop}. As in \eqref{eq:DeltaMinus1}, the result can be expanded in powers of $(\lambda_* n)^{2/3}$ and reads: 
\begin{equation}\label{eq:Delta1LoopLarge1}
\Delta_0  = \left[\alpha+\frac{5}{24}\log\left(\frac{\lambda_* n}{8\pi^2}\r)\right]\left(\frac{\lambda_* n}{8\pi^2}\r)^{4/3}+
\left[\beta-\frac{5}{36}\log\left(\frac{\lambda_* n}{8\pi^2}\r)
\right]\left(\frac{\lambda_* n}{8\pi^2}\r)^{2/3}+\mO(1),
\end{equation}
where the coefficients $\alpha$ and $\beta$ are
\begin{equation}
\alpha=-0.5753315(3),\qquad
\beta=-0.93715(9).
\end{equation}
The logarithmic terms are computed analytically, while the coefficients $\alpha$ and $\beta$ follow from a numerical fit. Details of the calculation are given in the appendix \ref{app:1loopCylLarge}.
The structure of the result \eqref{eq:Delta1LoopLarge1} is in agreement with the expected form of the large charge expansion in $d$ dimensions. This is evident summing \eqref{eq:Delta1LoopLarge1} to the leading order in \eqref{eq:DeltaMinus1} and writing the result in the form
\begin{multline}\label{eq:Delta1LoopLarge2}
\Delta_{\phi^n}=\frac{1}{\eps}\l(\frac{2}{5}\eps n\r)^{\frac{4-\eps}{3-\eps}}\l[
\frac{15}{8}+\eps \l(\alpha+\frac38\r)+\mO\l(\eps^2\r)\r]\\
+\frac{1}{\eps}\l(\frac{2}{5}\eps n\r)^{\frac{2-\eps}{3-\eps}}\l[
\frac{5}{4}+\eps \l(\beta-\frac{1}{4}\r)+\mO\l(\eps^2\r)\r]
+\mO\left((\eps n)^0\right).
\end{multline}
The change in the exponents of the $(\eps n)$ terms with respect to the leading order \eqref{eq:DeltaMinus1} account for the logarithms in \eqref{eq:Delta1LoopLarge1}.
Recalling that $d=4-\eps$, eq. \eqref{eq:Delta1LoopLarge2} is clearly in agreement with the structure predicted in \cite{Hellerman:2015nra,Monin:2016jmo}, which is:
\begin{equation}\label{eq:DeltaLargeGeneral}
\Delta_{n}=n^{\frac{d}{d-1}}\left[c_0(d)+c_1(d)n^{-\frac{2}{d-1}}
+c_2(d)n^{-\frac{4}{d-1}}
+\ldots\right]
+n^0\left[b_0(d)+b_1(d)n^{-\frac{2}{d-1}}+\ldots\right].
\end{equation}
From the point of view of the large charge EFT, the first term is a purely classical contribution, while the second term is the one-loop Casimir energy of the Goldstone mode\footnote{In non-even dimensions, this term is independent of the Wilson coefficients of the EFT and is hence universal \cite{Hellerman:2015nra}; for instance, $b_0(3)\simeq -0.937$.}.
 We have checked that the coefficients of the logarithms multiplied by subleading powers of $(\lambda_* n)$ ensure the agreement between our result and the predicted structure \eqref{eq:DeltaLargeGeneral} also in the subleading orders in $n$. The large $\lambda_* n$ expansion of the classical result determines the coefficients $c_i(d)$ at leading order, while eq. \eqref{eq:Delta1LoopLarge2} determines $c_0(d)$ and $c_1(d)$ to order $\mO\l(\eps\r)$. Even though we computed also the 
coefficient of the $(\lambda_* n)^0$ term in \eqref{eq:Delta1LoopLarge1} (see eq. \eqref{eq:AppFit}), in the expansion of \eqref{eq:DeltaLargeGeneral} for $d=4-\eps$ to first order, we cannot disentangle the first correction in $\eps$ to $c_2(d)$ and the leading order value of $b_0(d)$ (which is zero at tree-level). 

%

\section{Discussion}\label{secDiscussion}

\subsection{Large order behavior}

Expanding all functions $\D_\ell$ in a power series in $\lambda_* n$
\be
\Delta_\ell = \sum_{k} f_{\ell,k}  (\lambda_* n)^k,
\ee
it naively seems that the anomalous dimension (\ref{eq:cylinderDeltaExpansion}) has, at fixed order in the semiclassical expansion, contributions from arbitrarily large powers of $n$. This, however, does not match the  diagrammatic computation which is valid for small $\lambda_*n$ but virtually large $n$. Indeed, beyond  order $\lfloor n/2 \rfloor$ in the ordinary loop expansion the operator $\phi^n$ does not have enough free legs to provide  terms with higher and higher powers of $n$.

To understand what happens from the semiclassical perspective, we can compare contributions to the anomalous dimension that are of the same order in $\lambda_*$ but which come from different orders  in the semiclassical expansion.  For instance we can consider
$\D_\ell$ and $\D_{\ell+1}$. The contributions of the same order in $\lambda_*$ are controlled by $\lambda_*^{\ell+k} f_{\ell,k}n^k$ and $\lambda_*^{\ell+k} f_{\ell+1,k-1} n^{k-1} $ respectively. Therefore, if 
\be
\f{f_{\ell+1,k-1}}{f_{\ell,k}} \sim k, 
\ee there can be a potential cancellation at order $k\sim n$, thus resulting in the correct behavior of the anomalous dimensions for $k$ beyond roughly  $ \lfloor n/2 \rfloor$. We checked that this is precisely what happens for $f_{-1,k}$ and $f_{0,k-1}$.

\subsection{Boosting diagrammatic loop calculations}

At the Wilson-Fisher fixed point, the expansion in \eqref{gammastructure} for the anomalous dimension of $\phi^n$, valid for small $\varepsilon n$, is written as
\begin{equation}\label{eq:gammastructureWF}
\gamma_{\phi^n}=n\sum_{\ell=1}\eps^\ell P_\ell(n),
\end{equation}
Hence, at any fixed order $\ell$ in \eqref{eq:gammastructureWF} there are $\ell$ independent coefficients to be determined. We can thus take advantage of existing results in the literature, as well as of the small $\lambda_* n$ expansion of our results \eqref{eq:DeltaTree} and \eqref{eq:Delta1loop}, to fix some or all of them. The anomalous dimensions of $\phi$, $\phi^2$ and $\phi^4$ are known to order $\eps^5$ with analytical coefficients \cite{Kleinert:2001ax,Calabrese:2002bm}, while the anomalous dimension of $\phi^3$ is known to the same order with numerical coefficients \cite{DePrato:2003yd}. These results then provide four constraints on each of the first five orders in \eqref{eq:gammastructureWF} and are enough to fix all the coefficients in $P_1(n),P_2(n)$ and $P_3(n)$. Furthermore, expanding the results \eqref{eq:DeltaTree} and \eqref{eq:Delta1loop} derived in this paper to order $\mO(\eps^5 n^5)$, we have a total of six constraints on each of the first five orders in \eqref{eq:gammastructureWF}. This clearly fully fixes the form of the five polynomials $P_1(n),P_2(n),\ldots,P_5(n)$. The form of the first two was given in \eqref{eq:gammaTwoLoop}, while the others read
\begin{align}
P_3(n)=&\frac{n^3}{125}+\frac{n^2 \l[16 \zeta (3)-29\r]}{500} +\frac{n \l[599-672 \zeta (3)\r]}{5000}+\frac{ \l[1024 \zeta (3)-603\r]}{10000},\\ \nn
P_4(n)=&-\frac{21 n^4}{5000}+\frac{n^3 \l[214-77 \zeta (3)-80 \zeta (5)\r]}{5000}+\frac{n^2 \left[66336 \zeta (3)+160 \pi ^4-89491\right]}{600000}\\ 
&+\frac{n \left[41073-45864 \zeta (3)+46720 \zeta (5)-224 \pi ^4\right]}{200000}
\\ \nn
&+\frac{75888 \zeta (3)-130560 \zeta (5)+512 \pi ^4-53717}{600000},\\  \nn
P_5(n)=& \frac{n^5 8}{3125}+
\frac{n^4 \l[476 \zeta (3)+480 \zeta (5)+448 \zeta (7)-1683\r]}{50000}\\
&+0.00093 n^3-0.01067 n^2-0.2460 n+0.2680.
\end{align}
We checked that $P_3(n)$ agrees both with the previous literature and our results, providing another non trivial check of our approach. The polynomial $P_4(n)$ was determined using our results and those in the literature for $\phi,\phi^2$ and $\phi^4$; we checked that it agrees numerically within $10\%$ level with the coefficient reported in \cite{DePrato:2003yd} for $\phi^3$. We do not know if this discrepancy is due to the numerical uncertainty of this result, as the latter is not reported in \cite{DePrato:2003yd}. For the same reason, we cannot quote the uncertainty on the last four coefficients of $P_5(n)$.

\subsection{Comparison with Monte-Carlo results at large charge}

We can compare our result in the large $(\lambda_* n)$ limit, given by \eqref{eq:Delta1LoopLarge2} in the first two leading orders, with the recent results of Monte-Carlo lattice simulations of the three-dimensional $O(2)$ model \cite{Banerjee:2017fcx}.
There, the authors computed the scaling dimensions of the lightest charge $n$ operator for various values of $n$ and compared their result with the predicted form \eqref{eq:DeltaLargeGeneral}, which in $d=3$ reads:
\begin{equation}
\Delta_n\simeq c_{3/2}n^{3/2}+c_{1/2}n^{1/2}-0.0937+c_{-1/2}n^{-1/2}
+\mO\l(n^{-1}\r).
\end{equation}
The authors there determined the coefficients $c_{3/2}$ and $c_{1/2}$ fitting the result of the lattice computation.

\begin{table}[t!]
\renewcommand*{\arraystretch}{1}
\centering
\begin{tabular}{ c||c|c } 
 & $c_{3/2}$ & $c_{1/2}$ 
 \\ \hline \hline
Monte-Carlo \cite{Banerjee:2017fcx} & $0.337(3)$ & $0.27(4)$
 \\ \hline
$\eps$-expansion: LO & $0.47$ & $0.79$ 
\\ \hline
$\eps$-expansion: NLO & $0.42$ & $0.04$ 
\end{tabular}
\caption{Comparison of the Monte-Carlo result in \cite{Banerjee:2017fcx} with the $\eps$-expansion; we display both the leading order (LO) result as well as the next to leading order (NLO).}\label{tab:MC}
\end{table}

We compared the coefficients they obtained with those which follow from \eqref{eq:Delta1LoopLarge2} putting $\eps=1$. The results are displayed in the table \ref{tab:MC}. Using the next to leading order contribution as an estimate of the error, the result  for $c_{3/2}$ is roughly within two standard deviations from the Monte-Carlo result, while for $c_{1/2}$ the error is as big as the leading order, making  a quantitative analysis impossible. It is however interesting to notice that for both coefficients the next to leading order values are closer than the leading order ones to the results obtained by the Monte-Carlo. It would be interesting to compute the two-loop order result to explore the convergence properties of the expansion.

\section{Outlook}\label{secOutlook}

In this paper we illustrated a situation where amplitudes involving a large number $n$ of legs
can be reliably  computed through a systematic semiclassical expansion. That the large number of legs be related to a large conserved charge was essential to achieve our goal, but also the specialization to a conformally invariant fixed
point  made the task technically easier. The main results, obtained in the context of the $U(1)$ Wilson-Fisher
fixed point, were already illustrated in the introduction and we will not repeat them here. Instead we would here like to provide a perspective on  future research.

The most obvious extension  concerns the application of  our method to different models.
The $U(1)$ model with sextic interaction $\lambda \bar \phi^3\phi^3$ expanded around $d=3$ allows the most direct generalization. We have already significantly progressed in that study and a paper will appear shortly.
With respect to the quartic case, the most interesting novelty  of the sextic case is that the $\beta$-function of $\lambda$ arises at 2-loops, so that the model is conformally invariant up to 1-loop in exactly $d=3$. That allows to  compare the semiclassical  and diagrammatic computations at their respective 1-loop orders while working in exactly $d=3$. One thus obtains a nice diagrammatic check
of the Casimir contribution to the operator dimension at large charge derived using the universal superfluid description in \cite{Hellerman:2015nra,Bern1,Monin:2016jmo}. Another direction to explore, concerns Wilson-Fisher fixed points in  models with non-abelian
symmetry, like the $O(n)$ model. That will allow to study the patterns of symmetry breaking induced by the choice of the Cartan charges, again illuminating the more general, but abstract, work in refs \cite{Monin:2016jmo,HellermanO41,HellermanO42}

While most of the above are low hanging fruits, there are also  structural questions whose study is possibly more technically involved. One concerns the spectrum of nearby operators with the same charge.
For instance operators with two more additional derivatives like $\phi^{n-2}\partial_\mu\phi\partial^\mu\phi$ or $\phi^{n-2}\partial_\mu\phi\partial^\nu\phi$. These, along the lines drawn by the refs \cite{Hellerman:2015nra,Monin:2016jmo}, will be associated with the 
Fock space of excitations around the leading semiclassical solution. The  novelty with respect to refs
\cite{Hellerman:2015nra,Monin:2016jmo} is here the presence of the  parameter $\lambda_* n$, which controls the transition from the small to the large charge regime, and, relatedly, the presence of the mode associated to the radial variable $\rho$. In fact $\lambda n$ precisely controls the mass of this excitation and the transition to the pure superfluid regime $\lambda n\gg 1$ where it is superheavy. The resulting spectrum as a function of $\lambda n$ will thus provide a closed form
description of the transition, which we already mentioned resembles the small to large 't Hooft coupling transition in the AdS/CFT description of $N=4$ SYM.

Finally, another obvious, if perhaps technically involved problem, is the computation of 3-point functions, which would extend control to the full set of CFT data. In principle, given our control of the full theory, we should also be able to compute correlators of the form $\langle \bar \phi^{n_1+n_2}\phi^{n_1}\phi^{n_2}\rangle$ with $n_{1,2}$  large. The computation of this quantity would require to find the stationary solution in the presence of insertions of the terms $\log \rho$ in the path integral exponent. There should be no way to bypass this as we did in this paper for the computation of the operator dimension. The result may require a numerical analysis, but we have not yet investigated it. 

More  directions may later appear, but  the avenue indicated by our paper seems  promising already at this stage.

\subsection*{Acknowledgements}

We would like to thank Sergei Dubovsky, Anton de la Fuente, Joao Penedones, Slava Rychkov, Marco Serone and Sergey Sibiryakov for useful discussions. The work of G.B., G.C. and R.R. is partially supported by the Swiss National Science Foundation under contract 200020-169696 and through the National Center of Competence in
Research SwissMAP. The work of AM is supported by ERC-AdG-2015 grant 694896. R.R. acknowledges  KITP at U.C. Santa Barbara for hospitality and support during the completion of this work.

\appendices

\section{Diagrammatic two loop computation in \texorpdfstring{$\lambda|\phi|^4$}{lambda phi4} \label{app:2loopPlane}}

\vspace{-0.8cm}

In this section we compute the anomalous dimension of the $[\phi^n]$ operator to two loop via diagrammatic techniques.
For simplicity, we work in momentum space and we consider an insertion of the operator $\phi^n$ within $n$ equal incoming momenta $p$. 
We want to compute, according to the definitions (\ref{eq:bareRenormalized}),(\ref{eq:renormalizedPhiN2}):
\begin{equation}
  \label{eq:ZphiNbis}
  \langle \phi^n \bar{\phi}(p)\bar{\phi}(p)\dots\bar{\phi}(p)\rangle = Z_{\phi^n} Z_\phi^n \langle [\phi^n]\, [\bar{\phi}](p)\, [\bar{\phi}](p)\dots[\bar{\phi}](p)\rangle
\end{equation}
and find the right renormalization constant $Z_{\phi^n}$ such that $\langle [\phi^n]\, [\bar{\phi}](p)\, [\bar{\phi}](p)\dots[\bar{\phi}](p)\rangle$ is finite in the minimal subtraction (MS) scheme. At two loop $Z_\phi$ is 
\cite{Kleinert:2001ax}
\begin{equation}
  \label{eq:Zphi}
  Z_\phi = 1- \frac{\lambda^2}{(16\pi^2)^2 8 \varepsilon} + \mathcal O(\lambda^3).
\end{equation}

\begin{figure}[t]
\centering
 \subfloat[]{\label{fig:diagramA}\raisebox{0.5cm}{\includegraphics{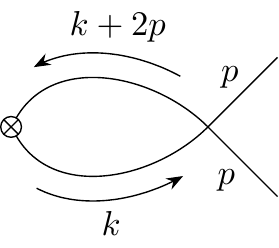}}} \hspace{2cm}
 \subfloat[]{\label{fig:diagramB}\raisebox{0.0cm}{\includegraphics{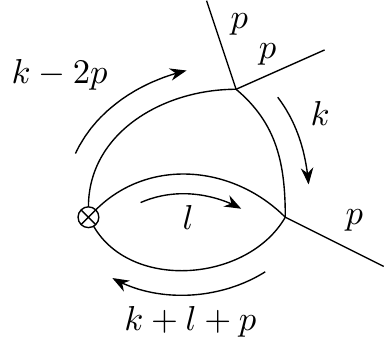}}} 
 
 \vspace{-0.5cm}
 \subfloat[]{\label{fig:diagramC}\raisebox{0.0cm}{\includegraphics{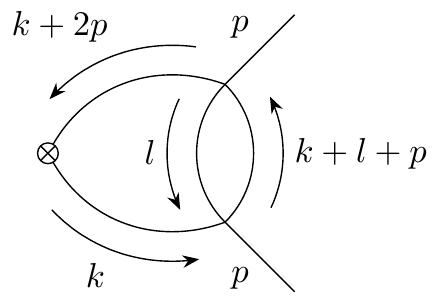}}} \hspace{1.2cm}
 \subfloat[]{\label{fig:diagramD}\raisebox{0.3cm}{\includegraphics{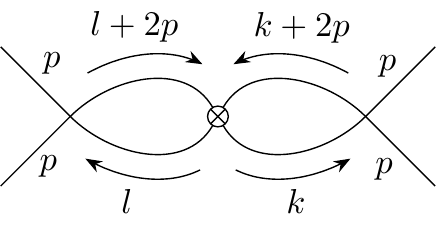}}} \hspace{1.2cm}
 \subfloat[]{\label{fig:diagramE}\raisebox{0.3cm}{\includegraphics{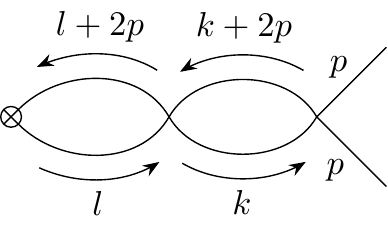}}}
 
 \vspace{-0.5cm}
 \hspace{-1.2cm}
 \subfloat[]{\label{fig:diagramG}\raisebox{0.0cm}{\includegraphics{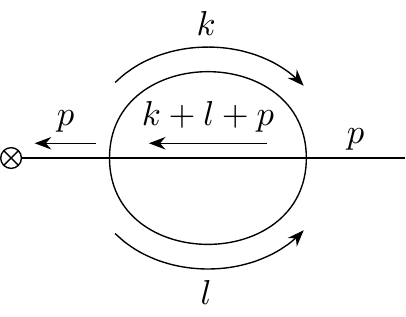}}} \hspace{1.8cm}
 \subfloat[]{\label{fig:diagramF}\raisebox{0.35cm}{\includegraphics{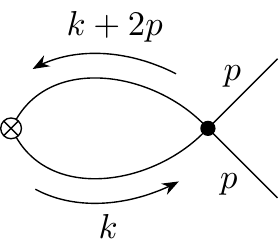}}} 
 
  \caption{\label{fig:diagrams} Feynman diagrams that contribute at two-loops.}
\end{figure}

We work within renormalized perturbation theory, the Feynman rules are:
\begin{equation}
  \btf{a}
    \vertex (a);
    \vertex [right=1.4cm of a] (b);
    \diagram*{(a) -- [momentum={[arrow distance=1.5mm]$p$}] (b),};
  \etf
  ~ = ~ \frac{1}{p^2}
  \hspace{50pt}
  \btf{x}
    \vertex (x);
    \vertex [above left=.6cm of x] (a);
    \vertex [above right=.6cm of x] (b);
    \vertex [below left=.6cm of x] (c);
    \vertex [below right=.6cm of x] (d);
    \diagram*[inline=(x)]{
      (a) --  (x),
      (b) --  (x),
      (x) --  (c),
      (x) --  (d),
    };
  \etf
   = ~  -\lambda
   \hspace{50pt}
   \btf{x}
    \vertex[bigdot] (x) {};
    \vertex [above left=.6cm of x] (a);
    \vertex [above right=.6cm of x] (b);
    \vertex [below left=.6cm of x] (c);
    \vertex [below right=.6cm of x] (d);
    \diagram*[inline=(x)]{
      (a) --  (x),
      (b) --  (x),
      (x) --  (c),
      (x) --  (d),
    };
  \etf
  = ~ -\delta_\lambda
\end{equation}
where $\delta_\lambda = \frac{5 \lambda^2}{16\pi^2\varepsilon}$ is the coupling counterterm at one loop in MS \cite{Kleinert:2001ax}. The $\phi^n$ operator will be represented by a crossed vertex and normalized to 
\begin{equation}\label{eq:diagram0}
  \feynmandiagram[small]{a [bigcross]}; = 1.
\end{equation}
All diagrams to two loop are displayed in figure \ref{fig:diagrams}. We don't represent the incoming lines if they are directly connected to the $\phi^n$ operator, only those connected to other vertices are shown.

The one loop diagram is:
\begin{equation}\label{eq:diagramA}
    \begin{split}
      \mathrm{\subref{fig:diagramA}} & = ~ \frac{n(n-1)}{2}\frac{1}{2}(-\lambda)
	\!\int\!\!\frac{\dd^d k}{(2 \pi)^d} \frac{1}{k^2} \frac{1}{(k+2p)^2} \\
      & = ~ -\frac{\lambda}{16\pi^2}\frac{ n(n-1)}{4}\left( \frac{2}{\varepsilon} + 2-\gamma+\log\left(\frac{\pi M^2}{p^2}\right)\right)+ \mathcal O(\varepsilon)
    \end{split}
\end{equation}
where in the first line, the first factor $\frac{n(n-1)}{2}$ indicates the number of ways the external momenta can be connected to form this diagram: one has to chose 2 momenta among $n$. The next factor $\frac{1}{2}$ is the usual symmetry factor, then comes the vertex, and finally the loop integral. In the result, $M$ is the scale introduced in (\ref{eq:bareRenormalized}).

Six diagrams have to be computed at two loop level. We need only the divergent piece of these diagrams. The procedure to compute the first two diagrams is described in \cite{Peskin:1995}. The last diagram includes the one loop counterterm~$\delta_\lambda$.
\begin{eqnarray}
   \mathrm{\subref{fig:diagramB}} & = & \frac{n(n-1)(n-2)}{2}\frac{1}{2}(-\lambda)^2
    \int\!\!\frac{\dd^d k}{(2\pi)^d} \int\!\!\frac{\dd^d l}{(2\pi)^d} \frac{1}{k^2}\frac{1}{(k-2p)^2}\frac{1}{l^2}\frac{1}{(k+l+p)^2} \nonumber \\
    & = & \frac{\lambda^2}{(16\pi^2)^2}\frac{n(n-1)(n-2)}{4}\left( \frac{2}{\varepsilon^2} + \frac{5-2\gamma + 2\log\left( \frac{\pi M^2}{p^2} \right)}{\varepsilon}\right) + \mathcal O(\varepsilon^0) \label{eq:diagramB}
\end{eqnarray}
\begin{eqnarray}
   \mathrm{\subref{fig:diagramC}} & = & \frac{n(n-1)}{2}(-\lambda)^2
    \int\!\!\frac{\dd^d k}{(2\pi)^d} \int\!\!\frac{\dd^d l}{(2\pi)^d} \frac{1}{k^2}\frac{1}{(k+2p)^2}\frac{1}{l^2}\frac{1}{(k+l+p)^2} \nonumber \\
    & = & \frac{\lambda^2}{(16\pi^2)^2}\frac{n(n-1)}{2}\left( \frac{2}{\varepsilon^2} + \frac{5-2\gamma + 2\log\left( \frac{\pi M^2}{p^2} \right)}{\varepsilon}\right) +  \mathcal O(\varepsilon^0) \label{eq:diagramC}
\end{eqnarray}
\begin{eqnarray}
   \hspace{-10pt}\mathrm{\subref{fig:diagramD}} & = & \frac{n(n-1)(n-2)(n-3)}{8}\frac{1}{4}(-\lambda)^2 
    \left( \int\!\!\frac{\dd^d k}{(2\pi)^d} \frac{1}{k^2}\frac{1}{(k+2p)^2}\right)^2 \nonumber \\
    & = & \frac{\lambda^2}{(16\pi^2)^2}\frac{n(n-1)(n-2)(n-3)}{8}\left( \frac{1}{\varepsilon^2} + \frac{2-\gamma + \log\left( \frac{\pi M^2}{p^2} \right)}{\varepsilon}\right) +  \mathcal O(\varepsilon^0) \label{eq:diagramD}
\end{eqnarray}
\begin{eqnarray}
   \mathrm{\subref{fig:diagramE}} & = & \frac{n(n-1)}{2}\frac{1}{4}(-\lambda)^2
    \left( \int\!\!\frac{\dd^d k}{(2\pi)^d} \frac{1}{k^2}\frac{1}{(k+2p)^2}\right)^2 \nonumber \\
    & = & \frac{\lambda^2}{(16\pi^2)^2}\frac{n(n-1)}{2}\left( \frac{1}{\varepsilon^2} + \frac{2-\gamma + \log\left( \frac{\pi M^2}{p^2} \right)}{\varepsilon}\right) +  \mathcal O(\varepsilon^0) \label{eq:diagramE}
\end{eqnarray}
\begin{eqnarray}
  \mathrm{\subref{fig:diagramG}} & = & n\frac{1}{2}(-\lambda)^2 \frac{1}{p^2}
  \int\!\!\frac{\dd^d k}{(2\pi)^d} \int\!\!\frac{\dd^d l}{(2\pi)^d} \frac{1}{k^2}\frac{1}{l^2}\frac{1}{(k+l+p)^2} \nonumber \\
  & = & -\frac{\lambda^2}{(16\pi^2)^2}\frac{n}{4\varepsilon} +  \mathcal O(\varepsilon^0) \label{eq:diagramG}
\end{eqnarray}
\begin{eqnarray}
   \mathrm{\subref{fig:diagramF}} & = & \frac{n(n-1)}{2}\frac{1}{2}(-\delta_\lambda)
    \!\int\!\!\frac{\dd^d k}{(2 \pi)^d} \frac{1}{k^2} \frac{1}{(k+2p)^2} \nonumber \\
    & = & -\frac{\lambda^2}{(16\pi^2)^2}\frac{5n(n-1)}{4}\left( \frac{2}{\varepsilon^2} + \frac{2-\gamma + \log\left( \frac{\pi M^2}{p^2} \right)}{\varepsilon}\right) +  \mathcal O(\varepsilon^0) \label{eq:diagramF}
\end{eqnarray}

Summing all contributions we get:
\begin{equation}
\begin{split}
  (\ref{eq:diagram0}) + & (\ref{eq:diagramA}) + (\ref{eq:diagramB}) + (\ref{eq:diagramC}) + (\ref{eq:diagramD}) + (\ref{eq:diagramE}) + (\ref{eq:diagramG})  + (\ref{eq:diagramF}) \\
  = \left( 1 - \frac{\lambda n (n-1)}{(16\pi^2) 2 \varepsilon} \right. & \left.+\ \frac{\lambda^2}{(16\pi^2)^2} \left(\frac{n^4-2n^3-9n^2+10n}{8\varepsilon^2}+ \frac{n^3-n^2-n}{4\varepsilon}\right) \right) \\
  & \times \left(1-\frac{\lambda n (n-1)\left(2-\gamma+\log\left(\frac{\pi M^2}{p^2}\right)\right)}{4 (16\pi^2)}\right) + \mathcal O(\varepsilon,\lambda^2\varepsilon^0) \\
\end{split}
\end{equation}
where the result, following (\ref{eq:ZphiNbis}), has been factored as $Z_{\phi^n} Z_\phi^n$, which contains only poles according to MS prescription, times the finite value of $\langle [\phi^n]\, [\bar{\phi}](p)\, [\bar{\phi}](p)\dots[\bar{\phi}](p)\rangle$.  This lets us compute the renormalization factor $Z_{\phi^n}$ using (\ref{eq:Zphi}):
$$
  Z_{\phi^n} = 1 - \frac{\lambda n (n-1)}{(16\pi^2) 2 \varepsilon} + \frac{\lambda^2}{(16\pi^2)^2} \left(\frac{n^4-2n^3-9n^2+10n}{8\varepsilon^2}+ \frac{2n^3-2n^2-n}{8\varepsilon}\right).
$$
The anomalous dimension $\gamma_{\phi^n}$ is computed using (\ref{eq:gammaPhiN}) and yields (\ref{eq:gammaTwoLoop}).

\section{Details of the one loop computation on the cylinder}

\subsection{Next to leading order corrections for generic \texorpdfstring{$\lambda n$}{lambda n}}\label{app:1loopCylsmall}

Here we discuss the derivation of \eqref{eq:Delta1loop} from \eqref{eq:Eminus1}.
To this aim, we first compute $\bar{e}_0$ expanding the first line in \eqref{eq:cylinderEnergyExpansion} from the expression of the bare coupling \eqref{eq:counterterm}:
\begin{multline}\label{eq:appBarE0}
\bar{e}_0(\lambda n,RM,d)=
e_0(\lambda n,d)+\left\{
\frac{5}{8}(\mu^2R^2-1)^2\left[\frac{1}{\eps}-\log(M\tilde{R})\right]\r.\\
\l.+\frac{1}{16 }(\mu ^2 R^2+3)( \mu ^2 R^2-1)+\mO(\eps)\right\}_{\lambda_0=\lambda},
\end{multline}
where we defined $\tilde{R}\equiv \sqrt{\pi } e^{\gamma/2} R$ and we used the equations of motion \eqref{eq:mu_f_equations} to expand the leading order in the coupling:
\begin{equation}
\frac{\p}{\p \lambda_0}\l[\frac{e_{-1}(\lambda_0 n,d)}{\lambda_0 R}\r]=\frac{R^{d-1}\Omega_{d-1}f^4}{16}.
\end{equation}
To compute $\Delta_0$ in \eqref{eq:cylinderDeltaExpansion}, we need to evaluate \eqref{eq:appBarE0} in $d=4$ and add the expansion of the leading order $\bar{e}_{-1}/\lambda$ to first order in $\eps$ (at fixed coupling)
\bea \nn
\Delta_0 &= &\left\{\bar{e}_0(\lambda n,RM,4)+\frac{\p}{\p\eps}\left[\frac{1}{\lambda }\bar{e}_{-1}(\lambda n,RM,4-\eps)\right]_{\eps=0}\right\}_{\lambda=\lambda_*}
\\
&= &\left\{
\lim_{\eps\rightarrow0}\left[\frac{R}{2}\sum_{\ell=0}^\infty n_{\ell}
\left[\omega_+(\ell)+\omega_-(\ell)\right]
+\frac{5}{8\eps}(\mu^2 R^2-1)^2\right]\right\}_{\lambda_0=\lambda_*}
\label{eq:AppIntermediate}
\eea
where the limit $\eps\rightarrow 0$ is taken at $\lambda_0$ fixed, we used eq. \eqref{eq:lambda} and 
\begin{equation}
\frac{1}{\lambda}\bar{e}_{-1}(\lambda n,RM,4-\eps)=
\frac{1}{\lambda M^\eps}e_{-1}(\lambda n M^\eps ,4-\eps).
\end{equation}
As anticipated, at the fixed point the dependence on the sliding scale drops.

To proceed, we need to isolate the divergent contribution in the sum in eq. \eqref{eq:AppIntermediate}. We use the $\ell\rightarrow\infty$ expansion of the summand
\begin{equation}\label{eq:AppOneLoopLexpansion}
n_{\ell}\left[\omega_+(\ell)+\omega_-(\ell)\right]\sim\sum_{n=1}^\infty
c_n\ell^{d-n}.
\end{equation}
The first five terms provide a divergent contribution in $d=4$. The expansion in $4-\eps$ dimensions of the coefficients is
\begin{equation}
\begin{gathered}
c_1=\frac{2}{R}+\mO\left(\eps\right),\quad
c_2=\frac{6}{R}+\mO\left(\eps\right),\quad
c_3=2 \mu ^2 R+\frac{4}{R}+\mO\left(\eps\right),\quad
c_4=2 \mu ^2 R+\mO\left(\eps\right),\\
c_5=-\frac{5 \left(\mu ^2 R^2-1\right)^2}{4 R}+\eps \frac{ \left[-225 \mu ^4 R^4+50 \mu ^2 R^2+150 \gamma  \left(\mu ^2 R^2-1\right)^2+113\right]}{120 R}
+\mO\l(\eps^2\r).
\end{gathered}
\end{equation}
We can now rewrite the sum isolating explicitly the divergent contribution as
\begin{equation}\label{eq:AppE1loop}
\frac{1}{2}\sum_{\ell=0}^\infty n_{\ell}
\left[\omega_+(\ell)+\omega_-(\ell)\right]=
\frac12 \sum_{n=1}^5c_n\sum_{\ell=1}^\infty \ell^{d-n}
+\frac12 \sum_{\ell=1}^\infty\bar{\sigma}(\ell)+\frac12\omega_+(0),
\end{equation}
where $\bar{\sigma}(\ell)$ is defined subtracting the first five terms in \eqref{eq:AppOneLoopLexpansion} from the original summand,
\begin{equation}
\bar{\sigma}(\ell)  = n_{\ell}\left[\omega_+(\ell)+\omega_-(\ell)\right]-\sum_{n=1}^5
c_n\ell^{d-n} ,
\end{equation}
and we used that $\omega_-(0)=0$. From \eqref{eq:AppOneLoopLexpansion} we see that the sum over $\bar{\sigma}(\ell)$ is convergent and can be evaluated directly in $d=4$. The first terms provide a divergent contribution which can be computed using $\sum_{\ell=1}^\infty \ell^{x}=\zeta(-x)$ and recalling $\zeta(1-\eps)\sim - 1/\eps$:
\begin{equation}\label{eq:AppE1loopDiv}
\frac12 \sum_{n=1}^5c_n\sum_{\ell=1}^\infty \ell^{d-n}=-
\frac{5 \left(\mu ^2 R^2-1\right)^2}{8  R\,\eps}-\frac{15 \mu ^4 R^4-6 \mu ^2 R^2+7}{16 R}.
\end{equation}
Using eq.s \eqref{eq:AppE1loop} and \eqref{eq:AppE1loopDiv} in \eqref{eq:AppIntermediate}, we obtain the result in the main text \eqref{eq:Delta1loop}.

\subsection{Next to leading order corrections for large \texorpdfstring{$\lambda n$}{lambda n}}\label{app:1loopCylLarge}

Here we discuss the derivation of the result \eqref{eq:Delta1LoopLarge1}. To this aim, it is convenient to start from eq. \eqref{eq:AppIntermediate}, derived in the previous appendix. We denote the summand in \eqref{eq:E1loop} with the bare coupling replaced by the renormalized one as
\begin{equation}
s(\ell,d)\equiv  n_\ell R\left[\omega_+(\ell)+\omega_-(\ell)\right]_{\lambda_0=\lambda}.
\end{equation} 
We then separate the sum over $s(\ell,d)$ into two terms introducing a cutoff $AR\mu$, where $A\gtrsim 1$ is an arbitrary number such that $AR\mu_*$ is an integer:
\begin{equation}
\frac12\sum_{\ell=0}^{\infty}s(\ell,d)=
\frac12\sum_{\ell=0}^{AR\mu}s(\ell,d)+\frac12\sum_{AR\mu+1}^{\infty}s(\ell,d).
\end{equation}
We can approximate the second sum using the Euler-Maclaurin formula:
\begin{equation}\label{eq:AppEulerMaclaurin}
\sum_{AR\mu+1}^{\infty}s(\ell,d)\simeq\int_{A R\mu}^\infty d\ell s(\ell,d)-
\frac{s(AR\mu,4)}{2}
-\sum_{k=1}^{N_1}\frac{B_{2k}}{(2k)!}s^{(2k+1)}(AR\mu,4)+\mO(\eps),
\end{equation}
where $B_{2k}$ are the Bernoulli numbers and $N_1$ is an integer.
As $s^{(k)}(AR\mu)\sim (AR\mu)^{1-k}$ and $\frac{B_{2k}}{(2k)!}$ approaches zero exponentially fast as $k$ grows, the error we make in \eqref{eq:AppEulerMaclaurin} can be made arbitrarily small increasing $N_1$.
The integral in \eqref{eq:AppEulerMaclaurin} is approximately evaluated using the expansion \eqref{eq:AppOneLoopLexpansion} truncated after $N_2$ terms, giving
\begin{equation}\label{eq:AppIntegral}
\begin{split}
\frac12\int_{A R\mu}^\infty d\ell s(\ell,d) &\simeq\frac{1}{2}(AR\mu)^{d}
\sum_{n=1}^{N_2}\frac{c_n}{(AR\mu)^{n-1}(n-1-d)} \\
&\equiv-\frac{5 \left(\mu ^2 R^2-1\right)^2}{8  \,\eps}+ \frac{5}{8} \left(R^2\mu ^2-1\right)^2 \log (R \mu )+f_{N_2,A}(R\mu)+\mO(\eps),
\end{split}
\end{equation}
where $f$ is a regular function of $R\mu$. As before, increasing $N_2$ we can improve at will the precision of our calculation for $A\gtrsim 1$. Using \eqref{eq:AppIntermediate} we then conclude
\begin{equation}\label{eq:AppDeltaLarge1}
\Delta_{0}=\frac{5}{8} \left(R^2\mu_*^2-1\right)^2 \log (R \mu_* )+
F(R\mu_*),
\end{equation}
where the function $F(R \mu_*)$ can be computed from
\begin{equation}\label{eq:AppNumericalObject}
F(R\mu_*)\simeq f_{N_2,A}(R\mu_*)-\frac{s(AR\mu_*)}{2}+
\left[\frac12\sum_{\ell=0}^{AR\mu_*}s(\ell,4)-\sum_{k=1}^{N_1}\frac{B_{2k}}{(2k)!}s^{(2k+1)}(AR\mu_*)\right]_{\mu=\mu_*}.
\end{equation}

The function $F(R\mu_*)$ can now be evaluated numerically and then fitted to the expected functional form, estimating the error from the first subleading terms neglected in the sums in \eqref{eq:AppEulerMaclaurin} and \eqref{eq:AppIntegral}. Using $N_1=4$, $N_2=10$ and $A=10$, we evaluated \eqref{eq:AppNumericalObject} for $R\mu_*=11,12,\ldots 210$. The result was fitted with an expansion in $(R\mu_*)^{-2}$, starting from $(R\mu_*)^4$, with four parameters\footnote{A fit with three parameter produces the same results with smaller standard errors.}. The first three terms read:
\begin{equation}\label{eq:AppFit}
F(R\mu_*)=-2.01444683(3)(R\mu_*)^4+2.49986(9)(R\mu_*)^2-0.55(4)+
\mO\left((R\mu_*)^{-2}\right).
\end{equation}
We have also verified that the coefficients of $(R\mu_*)$, $(R\mu_*)^3$, $(R\mu_*)^4\log(R\mu_*)$ and $(R\mu_*)^2\log(R\mu_*)$ are compatible with zero if included, individually or in combination, in the fit of the function in \eqref{eq:AppNumericalObject}. Notice that the functional form \eqref{eq:AppFit} agrees with \eqref{eq:DeltaLargeGeneral} for $d=4$ after expanding $R\mu_*$ in terms of $(\lambda_* n)^{2/3}$.

The expansion of the first term in \eqref{eq:AppDeltaLarge1} produces logarithms of $\lambda_* n$:
\begin{equation}\label{eq:AppLog}
\begin{split}
\frac{5}{8} \left(R^2\mu_*^2-1\right)^2 \log (R \mu_* ) =&
5 \left(\frac{(\lambda_* n)^{4/3} }{384 \pi ^{8/3}}
-\frac{(\lambda_* n)^{2/3}  }{144 \pi ^{4/3}}
+\frac{1 }{72}\right)\log \left(\frac{\lambda_* n}{8 \pi ^2}\right)\\
&+\frac{5}{288} \left(\frac{3 (\lambda_* n)^{2/3}}{\pi ^{4/3}}-10\right)+
\mO\left(\left(\frac{\lambda_* n }{16\pi^2}\right)^{-2/3}\right).
\end{split}
\end{equation}
As explained in the main text, the coefficients of the logarithms ensure that the one-loop result takes the form predicted by the large charge CFT predictions. Assuming that $F(R\mu_*)$ contains only powers of $R\mu_*$ (as we checked in \eqref{eq:AppFit}), one can verify that this is true for all the subleading orders in $(\lambda_* n)$ as well. Summing \eqref{eq:AppFit} and \eqref{eq:AppLog} and expanding $(R\mu_*)^2$ in powers of $(\lambda_* n)^{2/3}$, we obtain the result stated in the main text.

\bibliographystyle{JHEP}
\bibliography{epsilon}{}

\end{document}